\documentclass[reprint,amsmath,amssymb,prb,superscriptaddress]{revtex4-2}
\usepackage{graphicx}
\usepackage{xcolor}
\usepackage{soul}
\usepackage{newtxtext} 
\usepackage[margin=1in]{geometry} 
\usepackage{graphicx}
\usepackage{dcolumn}
\usepackage{hyperref}
\usepackage{amsmath}
\usepackage{cleveref}
\usepackage{newtxmath}

\hypersetup{colorlinks,
	linkcolor={blue!75!black!80!yellow},
	citecolor={blue!75!black!80!yellow},
	urlcolor={blue!75!black!80!yellow}
}

\makeatletter \renewcommand\@make@capt@title[2]{%
\@ifx@empty\float@link{\@firstofone}{\expandafter\href\expandafter{\float@link}}%
\sffamily{\textbf{#1}}\@caption@fignum@sep#2 }

\usepackage[normalem]{ulem}
\makeatletter
\renewcommand\@make@capt@title[2]{%
    \@ifx@empty\float@link{\@firstofone}{\expandafter\href\expandafter{\float@link}}%
    \sffamily{\textbf{#1}}\@caption@fignum@sep#2
}%

\begin{document}

\title{Non-Linear Arrhenius Behavior of Self-Diffusion in $\beta$-Ti and\ Mo}

\author{Yaxian Wang}
\thanks{These authors contributed equally.}
\affiliation{Department of Materials Science and Engineering, The Ohio State University, 2041 College Road, Columbus, OH 43210, USA}
\author{Zhangqi Chen}
\thanks{These authors contributed equally.}
\affiliation{Department of Materials Science and Engineering, The Ohio State University, 2041 College Road, Columbus, OH 43210, USA}
\author{Wolfgang Windl}
\email[Electronic address:\;]{windl.1@osu}
\affiliation{Department of Materials Science and Engineering, The Ohio State University, 2041 College Road, Columbus, OH 43210, USA}

\author{Ji-Cheng Zhao}
\email[Electronic address:\;]{jczhao@umd.edu}
\affiliation{Department of Materials Science and Engineering, University of Maryland, 4418 Stadium Drive, College Park, MD 20742, USA}

\date{\today}

\begin{abstract}
\noindent  While anomalous diffusion coefficients with non-Arrhenius like temperature dependence are observed in a number of metals, a conclusive comprehensive framework of explanation has not been brought forward to date. Here, we use first-principles calculations based on density functional theory to calculate self-diffusion coefficients in the bcc metals Mo and $\beta$-Ti by coupling quasiharmonic transition state theory and large displacement phonon calculations and show that anharmonicity from thermal expansion is the major reason for the anomalous temperature dependence. 
We use a modified Debye approach to quantify the thermal expansion over the entire temperature range and introduce a method to relax the vacancy structure in a mechanically unstable crystal such as $\beta$-Ti. 
Thermal expansion is found to weakly affect the activation enthalpy but has a strong effect on the prefactor of the diffusion coefficient, reproducing the non-linear, non-Arrhenius ``anomalous'' self-diffusion in both bcc systems with good agreement between calculation and experiment. 
The proposed methodology is general and simple enough to be applicable to other mechanically unstable crystals.
\end{abstract}

\maketitle


\section{\label{sec:intro}Introduction}

Self-diffusion in metals is believed to be understood rather well and their coefficients obey a linear temperature dependency on an Arrehenius plot (log($D$) vs. $1/T$). 
However, a number of elements show a remarkable ``curvature'' and thus deviation from linearity. 
While such an anomaly is rare for non-bcc metals, it has been found to be especially strong in the group IVB metals Ti, Zr and Hf, whose bcc-phases are mechanically unstable at lower temperatures, while it is much weaker in other bcc metals like Mo and Nb~\cite{neumann2011self} that are mechanically stable for all temperatures. 

Focusing on bcc Mo (in the following ``Mo'') and bcc Ti (in the following ``$\beta$-Ti'') as prototypes for mechanically stable and unstable bcc-metals, the underlying mechanism(s) for such a non-Arrhenius behavior has been under debate for many decades and remains unsettled.
Given that self-diffusion in elementary metals is governed by vacancy jumps, initial explanations were mostly based on the contribution of “secondary” diffusion mechanisms and include diffusion via divacancies~\cite{Fano1977vacancy}, next-nearest-neighbor (NNN) jumps~\cite{Ait-Salem1979investigation,goltz1980study}, diffusion via interstitials~\cite{schilling1978self} or diffusion enhancement by phase transformations~\cite{Sanchez1975model,Sanchez1978}. 
While none of these mechanisms could be conclusively confirmed for either Mo or  $\beta$-Ti, it has been argued in the late 1980s that secondary mechanisms would not be consistent with  experiments~\cite{Herzig1987,Petry1988atomistic}, and it was suggested that self-diffusion should be dominated by the traditional  mono-vacancy jumps through nearest-neighbor (NN) sites in both Mo and $\beta$-Ti~\cite{Petry1988atomistic,Vogl1989direct,Neumann1988}. 
This would leave some form of anharmonicity as the  explanation for the diffusion anomaly, since within harmonic transition state theory~\cite{rieth2007handbook,Stoddard20025abinitio}, the temperature dependence of the diffusion coefficient should exactly follow the Arrhenius equation $D(T) = D_0 {\rm exp}(-Q/k_BT)$ with temperature-independent prefactor $D_0$ and activation energy $Q$, which for mono-vacancy diffusion is the sum of vacancy formation enthalpy $E_f$ and migration enthalpy $E_m$. 
The anharmonicity was initially suggested to manifest itself in the form of soft phonons due to the specific distribution of the $d$-electrons ~\cite{kohler1987anomalous,kohler1988correlation}. 

%
By now, the continuous advances in atomistic simulation methods and computational capabilities have provided new opportunities to gain insight into the microscopic origin of anomalous diffusion, and lead to a variety of  explanations for non-Arrhenius diffusion. 
Proposed mechanisms include a vacancy-interstitial model in Ti~\cite{Smirnov2020non-arrhenius}, anharmonic effects in both $\beta$-Ti~\cite{Kadhodaei2020Phonon} and in Mo~\cite{Mattsson2009quantifying}, and concerted atomic motion~\cite{Sangiovanni2019superioniclike,Fransson2020}.

The previous work has been performed either with classical molecular dynamics (MD) based on empirical potentials or with density-functional theory (DFT).
On the classical MD side, in Ref.~\cite{Smirnov2020non-arrhenius} diffusion coefficients in $\beta$-Ti were determined from the mean-square displacement of the atoms.
The results suggested that interstitials would play a non-negligible role at self-diffusion near melting temperature and the sum of Arrhenius-like vacancy and interstitial diffusions with different slopes would result in the observed anomalous diffusion. 
The spontaneous formation of Frenkel pairs was also found in Ref.\ \cite{Fransson2020}, leading to  temperature-dependent formation energies. 
While no diffusion coefficients were calculated in Ref.~\cite{Fransson2020}, it was argued that the temperature-dependent formation energy would give rise to anomalous diffusion. 
The fact that the high-temperature defect concentrations varied by several orders of magnitude between Refs.~\cite{Smirnov2020non-arrhenius} and \cite{Fransson2020} as well as the fact that interstitial concentrations have been found exceedingly small in Mo even at its melting temperature casts doubt on this explanation. 
In another MD study \cite{Sangiovanni2019superioniclike}, it was proposed that concerted motion of atoms significantly contributed to diffusion at temperatures $>83$\% of the melting temperature, in addition to the otherwise dominant vacancy diffusion which followed a straight Arrhenius behavior over the entire examined temperature range from 73\% to 97\% of the melting temperature.
While the considerable variation in  results of these studies may at least partially originate from the different interatomic  potentials used~\cite{Ko2015,Hennig2008,dickel2018mechanical}, the fact that experiments show pronounced non-linearity already in the sub-1600~K range where all MD-studies find Arrhenius like behavior make further studies desirable. 

Other than  classical MD, density functional theory (DFT)-based methods do not rely on empirical interatomic potentials.
Performing harmonic transition-state theory based DFT calculations of diffusion coefficients started to become standard in the late 1990 for semiconductors~\cite{Windl1999} and were adopted into the metallic systems a decade later~\cite{wu2016high,angsten2014elemental,Mantina2008first,mantina2009first}.
Mattson \emph{et al.}~\cite{Mattsson2009quantifying} proposed a large quadratic temperature dependence of the vacancy formation energy in bcc Mo based on \emph{ab intio } molecular dynamics (AIMD) simulations, but did not study the temperature dependence of the formation entropy.
Further, the two data points obtained at $>2400$~K were insufficient to address the nonlinearity right above the transition temperature.
%
Non-MD based first principles calculations on the high temperature bcc phases of group IVB metals like titanium, zirconium and hafnium have remained quite challenging due to their mechanical instability at zero-temperature~\cite{wagner2008lattice,Antolin2012fast}, with a manifestation of decreasing energies and negative curvatures as a function of atomic displacement~\cite{Antolin2012fast} or strain~\cite{wagner2008lattice} in certain directions at 0~K.
Since phonon frequencies are determined by square roots of energy versus displacement curvatures, this corresponds to vibrational modes with imaginary frequencies. 
Since within the common harmonic or quasiharmonic approximations, vibrational free energies in solids are calculated from phonon frequencies, the appearance of imaginary frequencies  impedes free-energy calculations at finite temperatures for mechanically unstable crystals. 

A number of approaches have been suggested to overcome the limitations from mechanical instabilities through molecular-dynamics~\cite{Hellman2011lattice} or phonon calculations on cells with self-consistently displaced atoms~\cite{souvatzis2009self} to capture the temperature-induced anharmonicity, but most of them are computationally too expensive to perform calculations on supercells with point defects. 
Kadkhodaei \emph{et al.} employed a combination of self-consistent ab initio lattice dynamics (SCAILD) and the temperature dependent effective potential (TDEP) method and examined in a thorough study the influence of phonon anharmonicity on the diffusion coefficient~\cite{Kadhodaei2020Phonon} in $\beta$-Ti, but did not include thermal expansion and left the lattice parameters fixed. 
While the results could explain differences in magnitude between modeling and experiment that arise when straight harmonic transition state theory is used, it did not reproduce non-Arrhenius behavior, suggesting thermal expansion as a prime suspect to cause anomalous temperature dependence.


Herein, we perform for a comprehensive study of the thermal expansion and self-diffusion anomaly in the mechanically stable Mo and the unstable $\beta$-Ti with an accurate and efficient \emph{ab initio} approach.
Our work utilizes the large displacement method (LDM), first proposed by Antolin \emph{et al.}~\cite{Antolin2012fast}, from which we demonstrate that the thermal expansion coefficients are well captured in the temperature range where the quasi harmonic approximation (QHA) applies. 
For the minimum-volume at each temperature, we then calculate the free energy of vacancy formation and diffusion coefficients within LDM-based harmonic transition state theory, effectively forming a quasiharmonic version of it.
Our predicted self-diffusion coefficients and their anomalous temperature dependence in both Mo and $\beta$-Ti agree well with previous measurements, a strong indication that going beyond harmonic to quasiharmonic transition state theory explains anomalous temperature dependence of self-diffusion in a natural way that eliminates the need to invoke ad-hoc non traditional assumptions about the diffusion mechanism. 
Finally, we quantified the non-linearity in the self-diffusion coefficients and compared it with both experimental data and previous computational work.
We show that the non-linear Arrhenius self-diffusion arises mostly from the thermal expansion, and further pinpoint that it is not the migration enthalpy, but the formation entropy and, in the case of $\beta$-Ti, attempt frequency that carries the majority of the temperature dependence.
This work uncovers the nature of anomalous self-diffusion in Mo and $\beta$-Ti, and validates that LDM is an effective way to perform such calculations in mechanically unstable metals.


\section{\label{sec:method}Methodology}

\subsection{\label{subsec:method_TST}Self-Diffusion Theory for the bcc Lattice}
In the mono-vacancy mechanism of self-diffusion in the bcc lattice, the atom-vacancy exchange jump happens between two nearest neighbor positions along \textonehalf[111]. The bcc self-diffusion coefficient can be written as
\begin{equation}
\label{eq:diffusion_coefficient}
    D=gfa^2C_v\Gamma ,
\end{equation}
where $g$ is the geometrical factor (1 for cubic lattices), $f$ the correlation factor (0.7272 for bcc~\cite{Montet1973integral,compaan1956correlation}), $a$ the lattice constant, $C_v$ the vacancy concentration and $\Gamma$ the atom-vacancy exchange jump frequency along \textonehalf[111].

The atomic fraction of vacancy lattice sites $C_v$ is given by
\begin{equation}
\label{eq:vacancy_concentration}
    C_v = {\rm exp}\left(-\frac{\Delta G_f}{k_B T} \right)={\rm exp}\left(\frac{\Delta S}{k_B} \right)
    {\rm exp}\left(-\frac{\Delta H_f}{k_B T} \right),
\end{equation}
where $\Delta G_f$ is the free energy of vacancy formation within the bcc lattice. $\Delta S_f$ and $\Delta H_f$ are the vacancy formation entropy and enthalpy, respectively. The atom-vacancy exchange jump frequency $\Gamma$ can be calculated within transition state theory as~\cite{vineyard1957frequency}
\begin{equation}
\label{eq:jump_frequency}
    \Gamma = \nu^* {\rm exp} \left( -\frac{\Delta H_m}{k_B T}\right),
\end{equation}
where $\Delta H_m$ denotes the migration enthalpy and $\nu^*$ the attempt frequency. Substituting Eqs.~\ref{eq:vacancy_concentration} and \ref{eq:jump_frequency} into Eq.~\ref{eq:diffusion_coefficient}, the self-diffusion coefficients in bcc metals can be calculated by 
\begin{equation}
\label{eq:D_total}
D(T)=\overbrace{0.7272a^2\nu^*{\rm exp}\left(\frac{\Delta S_f}{k_B} \right)}^{D_0}{\rm exp}\left( -\frac{\overbrace{\Delta H_f + \Delta H_m}^Q}{k_B T}\right).
\end{equation}

Therefore, 4 parameters will be calculated to obtain self-diffusion coefficients through DFT calculations: $\Delta H_f$, $\Delta S_f$, $\Delta H_m$, and $\nu^*$.
Of those, $\Delta H_f$  and $\Delta H_m$ contribute to the thermal activation energy $Q$, while $\Delta S_f$ and $\nu^*$ contribute to the prefactor $D_0$. 
We will describe their calculation in Sec.~\ref{subsec:method_calculation_D}. 
Before that, the vacancy structures of Mo and $\beta$-Ti need to be relaxed, which is a straightforward task for Mo, but requires some extra attention for Ti, which we will discuss in Sec.~\ref{subsec:method_relaxation}. 
The basic parameters for the DFT calculations are then discussed in Appendix~\ref{Appendix:computational}, the large-displacement method in Sec.~\ref{subsec:method_LDM}, and the calculation of lattice expansion to introduce vibrational anharmonicity in Sec.~\ref{subsec:method_thermal_expansion}.

\subsection{\label{subsec:method_calculation_D}Calculation of the Diffusion Coefficient}

Formation quantities such as $\Delta G_f$, $\Delta S_f$ or $\Delta H_f$ can be calculated by the respective difference between perfect supercell and the relaxed cell with one vacancy. 
As an example, 
\begin{equation}
\label{eq:supercell_formation}
    \Delta G_f=G(X_{N-1}V_1)-\frac{N-1}{N}G(X_N),
\end{equation}
where $X$ is the lattice atom, $V$ indicates a vacancy and $N$ indicates the number of lattice sites in the supercell. 

In order to calculate the vibrational contribution to the free energies of formation at finite temperatures, $F_f(T)$, and the entropy of formation, $S_f(T)$, we use the quasi-harmonic approximation (QHA)~\cite{Baroni2001RMP} based on supercell $\Gamma$-point calculations as described in previous work~\cite{luo2009first}. 
There, the entropy is approximated by
\begin{eqnarray}
\label{eq:entropy}
    S(V,K) = k \Bigg\{ &\int& \frac{\frac{h\nu}{k_BT}}{{\rm exp}(\frac{h\nu}{k_BT})-1} g[\nu(V)] {\rm d}\nu \\ \nonumber
    &-&\int g[\nu(V)]{\rm ln}\left[1-{\rm exp}\left(-\frac{h\nu}{k_BT}\right)\right]{\rm d}\nu\Bigg\} ,
\end{eqnarray}
where $ g[\nu(V)]$ is the phonon density of states~\cite{luo2009first} and the $\Gamma$-point phonon frequencies $\nu(V)$ are calculated for a supercell with volume $V$.

The attempt frequency $\nu^*$ can be calculated as
\begin{equation}
    \label{eq:attemp_freq}
    \nu^* = \prod_{i=1}^{3N-3}\nu_i \bigg / \prod_{j=1}^{3N-4}\nu_j,
\end{equation}
where $\nu_i$ and $\nu_j$ are the normal mode ($\Gamma$--point phonon) frequencies of the stable and saddle point configurations, respectively, for a system of $N$ atoms and one vacancy. 
The product in the denominator specifically excludes the (imaginary) frequency corresponding to the unstable mode at the transition state. 

The climbing image nudged elastic band method (CI-NEB)~\cite{henkelman2000climbing} with 3 images was employed to determine the transition state (saddle point) structures for Mo and Ta (see Sec.~\ref{subsec:method_relaxation} why Ta), with the saddle point configuration of Ti determined from those as described in the following section due to the relaxation problems in mechanically unstable high-temperature structures. 
The initial and final structures, i.e. stable states, were fully relaxed first. Then during the CI-NEB calculations, all supercell volumes were fixed.

\subsection{\label{subsec:method_LDM}Large-Displacement Quasiharmonic Transition State Theory}
\begin{figure}
    \centering
    \includegraphics[width=\linewidth]{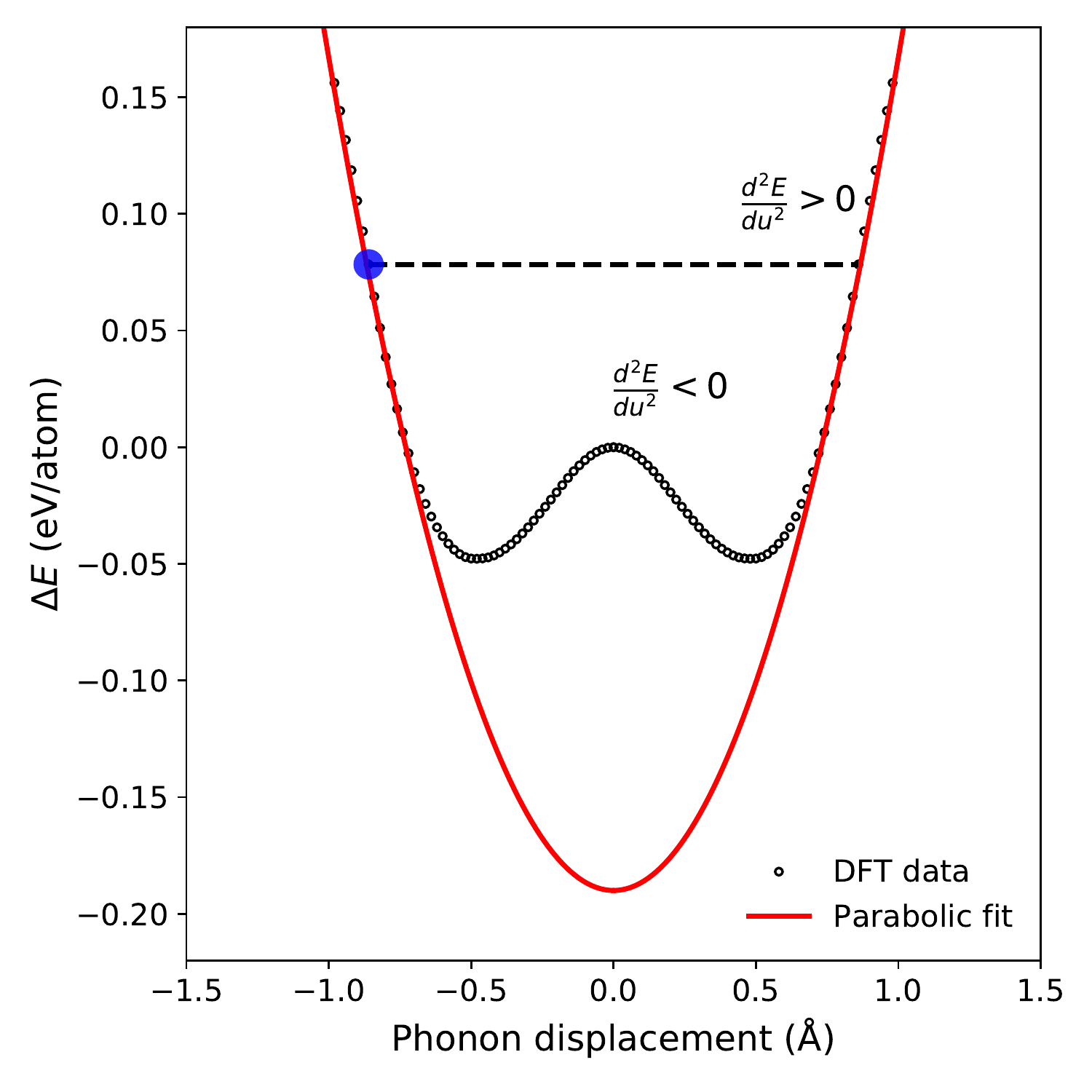}
    \caption{The energy difference (eV per Ti atom) with respect to the perfect cell in a  frozen-phonon cell for the unstable  $N$-point displacement pattern as a function of displacement (\AA) (black circles). 
    The energy ``hump'' at zero displacement highlights the mechanical instability, i.e. a negative curvature, or the second derivative of energy on phonon displacement ($\frac{d^2E}{du^2}<0$), of $\beta$-Ti at zero temperature.
    A harmonic potential well (red curve) is obtained by matching energy and curvature of the energy data at a large displacement value of 0.88~{\AA} as determined in Ref.~\cite{Antolin2012fast}.
    This represents the high temperature regime (black dashed line), where the atomic vibration exhibits large amplitude (blue marker) with high thermal energy, resulting in the envelope harmonic potential governing its vibration and real phonon frequencies ($\frac{d^2E}{du^2}>0$).
    }
    \label{fig:LD}
\end{figure}

The large displacement method (LDM) treats the anharmonicity through large atomic displacements into the high-temperature anharmonic range of the atoms’ energy wells. 
Within LDM, phonons are calculated from the curvature of the harmonic envelope of the quartic energy-vs.-displacement curve (red curve in \cref{fig:LD}).
At high temperatures, or above the transition temperature, the atoms actually sample at the large vibrational amplitudes, which makes them carry across the ``hump'' at zero displacement (black circles in \cref{fig:LD}), i.e. the local energetic barrier, and thus overcomes the mechanical instability of the bcc phase, therefore eliminating all imaginary frequencies from the phonon dispersions. 
Technically, this approach thus works like any standard finite-difference phonon calculation with the exception that the displacements are one or two orders of magnitude larger. 
The only question there is how large the displacement should be. In \cite{Antolin2012fast}, a large displacement value of 0.88~{\AA} was proposed for $\beta$-Ti, resulting in good agreement of theoretical and experimental phonon dispersion. 
The sensibility of this large-displacement value for $\beta$-Ti proposed is further confirmed by the comparison of the theoretical thermal expansion coefficients with the experimental values (\cref{subsec:results_thermal_expansion}). 
As was also shown in Ref.~\citenum{Antolin2012fast}, regular moderately small displacements  (we use 0.05~{\AA}) are the most sensible choice for Mo, which is mechanically stable at zero temperature.

\subsection{\label{subsec:method_relaxation}Relaxation of Vacancy Structure for Ti}
Due to $\beta$-Ti’s mechanical instability, one cannot simply relax its supercell with one vacancy at zero temperature like is commonly done for other crystals to determine the ground state structure. 
If one tried, one would end up with unphysically large atomic displacements and a negative formation energy as shown in result Sec.~\ref{subsec:results_Ti_relaxaion}. 
Because of the same reason, the CI-NEB method also cannot be used to find the saddle point structure. 
To circumvent this problem, we first perform an approximate, nearest-neighbor (NN) only, relaxation based on free energies for Ti, where free energy calculations based on the LDM (Sec.~\ref{subsec:method_LDM}) at three different temperatures between phase transformation temperature (1150~K) and melting temperature (1940~K) were performed for differently displaced NN shells in an otherwise unrelaxed vacancy cell. 
The minimum-energy displacement was then linearly extrapolated to zero temperature, $d_{\rm Ti}(T)=d_{\rm Ti}^0+\vartheta T$.
Since in the final calculations, we want to use full instead of NN only relaxation, we now use the zero-temperature extrapolated value to identify an appropriate surrogate structure for Ti from other mechanically stable bcc materials, scaled to the correct lattice constant. 
For NN-only relaxation, Mo is found to have one of the smallest, and Ta one of the largest relative NN relaxations. 
Thus, the positions of the atoms in the interpolated surrogate Ti-cell can be calculated from
\begin{equation}
    \label{eq:ti_position}
    x_{\rm Ti}=\frac{d_{\rm Ta}^0-d^0_{\rm Ti}}{d_{\rm Ta}^0-d^0_{\rm Mo}}x_{\rm Mo}+\frac{d_{\rm Ti}^0-d^0_{\rm Mo}}{d_{\rm Ta}^0-d^0_{\rm Mo}}x_{\rm Ta},
\end{equation}
a procedure that is easily transferable to other mechanically unstable bcc crystals. 
The saddle point structure of $\beta$-Ti is calculated in the same way from CI-NEB results for Mo and Ta. 
Results for these relaxation calculations are shown in Sec.~\ref{subsec:results_Ti_relaxaion}.

\subsection{\label{subsec:method_thermal_expansion}Thermal Expansion }
In the quasiharmonic version of transition state theory we employ in this paper, the four variables that determine vacancy self-diffusion according to Eq.~\ref{eq:D_total} are $\Delta H_f$, $\Delta S_f$, $\Delta H_m$ and $\nu^*$.
They are evaluated at the minimum-lattice constant in the Helmholtz free energy for each temperature, determined in turn from the traditional quasi-harmonic approximation to vibrational entropy. 
The Helmholtz free energy of a system as a function of volume ($V$) and temperature ($T$) is expressed as
\begin{eqnarray}
    \label{eq:helmholtz_energy}
    F(V,K) =E_{\rm tot}(V)&+& k_B T \Bigg\{  g[\nu(V)]h\nu {\rm d}\nu \\ \nonumber
    &+&\int g[\nu(V)]{\rm ln}\left[1-{\rm exp}\left(-\frac{h\nu}{k_BT}\right)\right]{\rm d}\nu\Bigg\} ,
\end{eqnarray}
where $E$ is the total (internal) energy at 0~K, in addition to the quantities defined for Eq.~\ref{eq:entropy}. 
As described in Sec.~\ref{subsec:method_LDM}, we used the LDM for the Ti phonon calculations and traditional displacements for Mo.

In order to determine the thermal expansion coefficient, we fit the lattice constant data vs. temperature with a modified expression derived from the Debye model~\cite{sayetat1998easy},
\begin{equation}
\label{eq:debye}
    a(T)=a_0[1+I_a(T)T\varphi(\theta_D/T)],
\end{equation}
where $a_0$ is the lattice parameter at 0~K, $\theta_D$ is the Debye temperature, 
and $\varphi$ is the Thacher function to approximate the Bose-Einstein weighted integral from the Debye specific heat, Eq.~(6) from ~\cite{thacher1960rational}.
Other than in the original model, we give $I_a$ a linear temperature dependence, $I_a(T)=I_0(1+\chi T)$ with $I_0$ and $\chi$ as fitting parameters in addition to $\theta_D$, since $I_a$, originally given by $I_a = K\gamma k_B/V_0$ ($K$ is the compressibility; $\gamma$ the Grüneisen constant; and $V_0$ the minimum volume of the cell) should be temperature dependent because both Gr\"uneisen parameter and elastic constants \cite{dickinson1967temperature} are temperature dependent.  
Without this, the typically observed increase in thermal expansion with temperature in the high-temperature limit is not described well enough. {The linear thermal expansion coefficient ($\alpha$) is finally calculated by $\alpha=\frac{1}{a_0}\frac{\partial a}{\partial T}$ from the temperature derivative of Eq.~\ref{eq:debye}.} 
\subsection{\label{subsec:method_non_linear}Quantification of Non-Linearity}
In order to quantify the non-linearity in our calculated and previous (experimental) data, we fit the logarithm of the diffusion coefficient in an Arrhenius plot with an equation analogous to bowing parameter dependent quadratic equations used for lattice parameters or elastic constants~\cite{Windl1998theory},
\begin{equation}
\label{eq:bowing}
    {\rm ln}D(\beta)=\frac{\beta-\beta_l}{\beta_h-\beta_l}{\rm ln}D_h
    +\frac{\beta_h-\beta}{\beta_h-\beta_l}{\rm ln}D_l
    +\eta \frac{(\beta-\beta_l)(\beta_h-\beta)}{(\beta_h - \beta_l)^2}.
\end{equation}

\color{black}
There, the first two terms are the linear interpolation between the logarithms of highest ($l$) and lowest ($l$) diffusion coefficients (respectively, $D_h$ and $D_l$ can be fitting coefficients), $\beta = 1/k_BT$, and $\eta$ is the ``bowing parameter'', i.e. the measure of the quadratic term, whose form is chosen to be zero at the extremal temperatures.
An absolute non-linearity parameter $\lambda$ can then be used to compare the degree of non-linearity between the different elements independent of if the temperature ranges exactly overlap or not. Following Eqs.~\ref{eq:jump_frequency} and \ref{eq:helmholtz_energy} in Ref.~\onlinecite{emancipator1993quantitative}, we calculate $\lambda$ by 
\begin{equation}
\label{eq:non_linearity}
    \lambda = \left\vert \frac{\eta}{\sqrt{30}({\rm ln}D_h - {\rm ln}D_l)}\right\vert.
\end{equation}

\subsection{\label{subsec:method_procedure}Overall Procedure}
In order to determine the diffusion coefficients for Ti and Mo within the quasiharmonic transition state theory employed in this paper, we performed the following steps: 	

(1)	The thermal expansion, i.e. the temperature dependence of the lattice constants, was calculated within the quasiharmonic approximation as described in Sec.~\ref{subsec:method_thermal_expansion} based on perfect bcc supercells for Ti and Mo.

(2) Supercells with one vacancy (initial and final position, separated by one bond length) were relaxed to their ground state in the traditional way for Mo and Ta, while the structure for Ti was determined as described in Sec.~\ref{subsec:method_relaxation}.
Then, the CI-NEB (Sec.~\ref{subsec:method_calculation_D}) was used to determine the saddle point configuration for Mo and Ta, while Ti was again gained from the interpolation procedure. 

(3)	The lattice constants calculated in (1) were assigned to the ground state and transition state (saddle point) structures for Mo and Ti determined in (2).

(4)	$\Gamma$-point phonon calculations using a finite difference method with displacements as described in Sec.~\ref{subsec:method_LDM} were performed for all temperature dependent structures, along with zero-temperature total energies. 

(5)	Self-diffusion coefficients at different temperatures were calculated based on the results from (4) using Eqs.~\ref{eq:D_total}-\ref{eq:attemp_freq}. 

(6) Analysis of the source of the non-linearity of the diffusion coefficient was carried out by quantifying the non-linearity of its different components as defined in Eq.~\ref{eq:D_total}. 
\subsection{\label{Appendix:computational}Computational details}
First-principles calculations were employed to perform all structural relaxations and CI-NEB runs as well as energy and phonon-frequency calculations. 
To describe ion-electron interactions, we used projector augmented plane-wave potentials~\cite{kresse1999ultrasoft} as implemented in the Vienna \emph{ab initio} simulation package (VASP) version 5.4.1~\cite{kresse1993ab,kresse1999ultrasoft}. 
Exchange and correlation contributions to the total energy were described by the Perdew-Burke-Ernzerhof (PBE) functional~\cite{Perdew1996generalized}. 
After convergence tests, we chose a 54-atom supercell ($3\times3\times3$ conventional bcc unit cells) with $4\times4\times4$ Monkhorst-Pack k-point mesh for Brillouin zone integration. 
Cutoff energies of 300~eV for Mo and 700~eV for Ti were found to be necessary for convergence in the phonon calculations. 
For Ta, the cutoff energy was 300~eV as well. The settings were kept constant for all runs.

\section{\label{sec:results}Results and Discussion}
\subsection{\label{subsec:results_thermal_expansion}Thermal Expansion in Mo and $\beta$-Ti }
\begin{figure*}
    \centering
    \includegraphics[width=\linewidth]{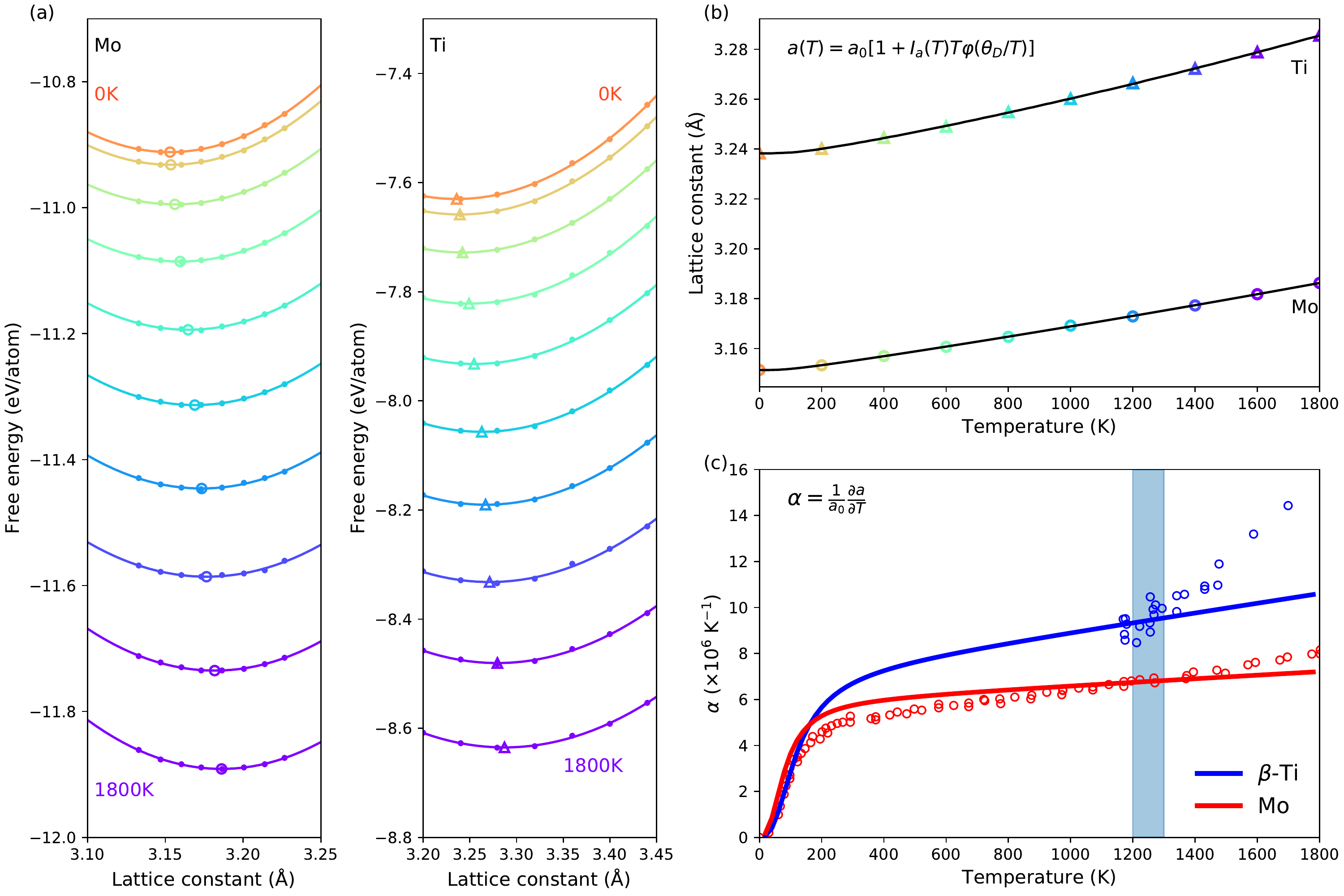}
    \caption{(a) Free energy for bcc Mo and bcc Ti with respect to lattice constant at temperatures from 0 to 1800~K with 200~K step are depicted by circles. The values are fit by quadratic curves (solid lines). 
    Crosses denote the energy minima of the respective curves and simultaneously the equilibrium lattice constant.
    (b) Ground-state lattice constants as a function of temperature for (a) Mo and (b) $\beta$-Ti (open circles), determined from the minima of the parabolic fits to DFT calculations of the Helmholtz free energy in~\cref{fig:fig1_thermal_expansion}a.
    (c) Calculated thermal expansion for $\beta$-Ti (blue, upper line and data) and Mo (red, lower line and data). 
    The lines are determined by fitting the DFT data from \cref{fig:fig1_thermal_expansion}b with Eq.~\ref{eq:helmholtz_energy} and taking the temperature derivative. 
    The circles denote experimental data from Ref.~\onlinecite{wang1998role} for Mo and Ref.~\onlinecite{touloukian1975thermophysical} for $\beta$-Ti.}
    \label{fig:fig1_thermal_expansion}
\end{figure*}

The results from the quasiharmonic approximation for the Helmholtz free energy vs. volume for temperatures between 0 and 1800 K in steps of 200 K for Mo and $\beta$-Ti are shown in Fig.~\ref{fig:fig1_thermal_expansion}(a). 
The energy minima are determined by a fit based on the assumption of a harmonic dependence of energy on the volume and are indicated by crosses in the plot at the minima of the parabolas. 
The minimum-energy lattice constants for both materials as a function of temperature are shown in Fig.~\ref{fig:fig1_thermal_expansion}(b).
The results for thermal expansion in Mo, based on traditional phonon calculations, are shown in Fig.~\ref{fig:fig1_thermal_expansion}(a). 
Fitting Eq.~\ref{eq:debye} to the DFT data results in parameter values of $a_0 = 3.1514 \AA$, $\theta_D$ = 327~K, $I_a = 5.91\times10^{-6} {\rm K}^{-1}$, and  $\chi = 6.42\times 10^{-6} {\rm K}^{-1}$. 
The lattice constant at 298~K is calculated with Eq.~\ref{eq:debye} to be 3.155~{\AA}, while the experimental lattice constant at room temperature is 3.147 ~{\AA}. 
The thermal expansion coefficient of Mo is calculated from the temperature derivative of Eq.~\ref{eq:debye} and is shown in Fig.~\ref{fig:fig1_thermal_expansion}c in comparison to experimental data from Ref.~\onlinecite{wang1998role}. 
While the (weak) increase with temperature in the calculations is slightly lower than what is found in experiment, the general agreement is rater good with an average deviation in the linear regime between 600 and 1800~K of ~5\%. 

The results for $\beta$-Ti, which are based on phonon calculations from the LDM, are shown in Fig.~\ref{fig:fig1_thermal_expansion}(a) and (b). 
Since, as described in Sec.~\ref{subsec:method_LDM}, the LDM uses the harmonic envelope of the quartic energy-vs.-displacement curve irrespective of if the kinetic energy of the vibrating atoms is high enough to overcome the energy ``hump'' at zero displacement, which in the real system only happens above the transition temperature, it can also be used to calculate the energy vs. volume dependence of metastable $\beta$-Ti below the transition temperature and thus is shown here for the entire range. 
As seen in Fig.~\ref{fig:fig1_thermal_expansion}(a), the calculated values are well described by a quadratic fit for all temperatures, as was the case for Mo. 
Thus, we can use the entire range for fitting Eq.~\ref{eq:helmholtz_energy} to get more reliable coefficients and find parameters of $a_0 = 3.2383\AA$, $\theta_D = 474$~K, $I_a = 7.08\times10^{-6} {\rm K}^{-1}$, and $\chi = 1.47\times10^{-4} {\rm K}^{-1}$. 

Experimental values compiled in Ref.~\onlinecite{touloukian1975thermophysical} are for pure Ti only available above the transition temperature from $\alpha$-Ti to $\beta$-Ti, which we have calculated previously to be at 1200~K with the LDM~\cite{Antolin2012fast}, within 4\% of the experimental value of 1155~K. 
The calculated lattice constant of $\beta$-Ti phase at 1200~K is 3.27~{\AA}, compared to an experimental value at the same temperature of 3.33~{\AA}. 
The calculated linear thermal expansion coefficient shows excellent agreement with the experimental values between 1200 and 1300~K with a considerable temperature dependence of $d\alpha /dT = 2\times 10^{-10} {\rm K}^{-2}$ (Fig.~\ref{fig:fig1_thermal_expansion}c). 
For higher temperatures, it seems that anharmonic effect other than large displacements become important as already discussed in Ref.~\cite{Antolin2012fast}, causing a faster increase in thermal expansion in experiment than captured by LDM alone. Therefore, we restrict our calculations here to this temperature range.

Overall, we have successfully determined the temperature dependence of the lattice constants in good agreement with experiments, which we will use in the following as input for vacancy and diffusion-coefficient calculations.

\subsection{\label{subsec:results_Ti_relaxaion}Groundstate of Vacancy Structure in $\beta$-Ti}
\begin{figure}
    \centering
    \includegraphics[width=\linewidth]{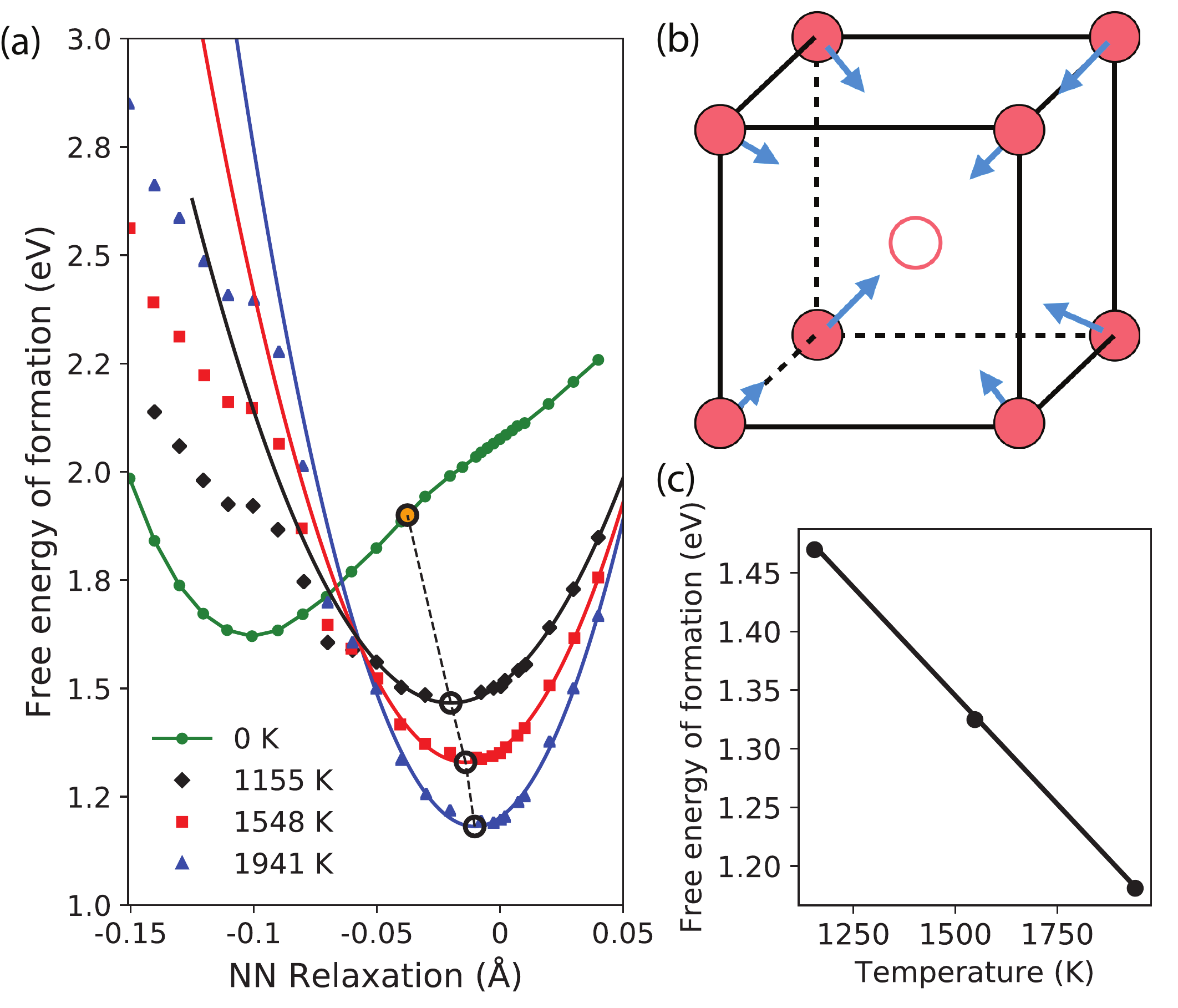}
    \caption{(a) Free-energy of formation for a vacancy in a 54-atom supercell of $\beta$-Ti as a function of simultaneous nearest-neighbor relaxation along (illustrated in b) the \{111\} bond directions with other atoms in fixed perfect lattice sites.
    Calculations were performed with DFT within the LDM quasiharmonic approximation. 
    The symbols are calculation results for 1155~K (black diamonds, experimental transition temperature), 1941~K (blue triangles, experimental melting temperature), and 1548~K (red squares, halfway in-between) with corresponding parabolic fits to find the minimum-energy displacements. 
    The green circles are zero-temperature results, and the orange dot the linear extrapolation vs. temperature from the high-temperature minimum-energy displacements (hollow black circles along the dashed line) to the zero-temperature curve. 
    (b) The nearest-neighbor displacement directions, with negative values moving towards the site of the vacancy (empty sphere).
    (c) Minimum-displacement free-energy of formation (indicated by the hollow circles in Figure 4) for a vacancy in a 54-atom supercell of $\beta$-Ti as a function of temperature (circles), along with a linear fit with  $F(T) = 1.90 {\rm eV} - 4.3k_BT$.}
    \label{fig:fig4_NNrelax}
\end{figure}

Following the procedure described in Sec.~\ref{subsec:method_relaxation}, we have performed free-energy calculations of vacancy formation energies in constant-volume cells of $\beta$-Ti (lattice constant the zero-temperature value from Fig.~\ref{fig:fig1_thermal_expansion}(b), 3.237~{\AA}) for three different temperatures (1155~K (experimental transition temperature), 1941~K (experimental melting point) and 1548~K (halfway in-between) as a function of simultaneous nearest-neighbor displacement along the bond direction to the vacancy. 
For each temperature, the minimum-energy displacement and energy minimum were determined from a parabolic fit. 
The results are shown in Fig.~\ref{fig:fig4_NNrelax}.

Fig.~\ref{fig:fig4_NNrelax}c shows the free energy of formation as a function of temperature for the three calculation temperatures. 
In principle, the calculations performed here result in the Helmholtz free energy of formation at constant volume. 
In order to estimate the energy error with respect to the Gibbs free energy from the non-zero pressure that comes from neglecting the formation volume of the vacancy [20], we estimate the formation volume from the pressure in the VASP output for the NN-displaced cell and the bulk modulus and pressure derivative reported previously~\cite{Antolin2012fast} through the pressure expression of the Murnaghan equation~\cite{Murnaghan1944}. 
We find typical pressures of $< 10$~kbar, indicating a volume relaxation in the supercell by less than 1\%. 
Multiplying the formation volume from that with the pressures, we find that the $pV$ term contributes less than 0.04~eV to the free energy, and thus is small. 
Fitting $F(T) = E_f -TS$ to the three values, we find $E_f = 1.90$~eV and $S_f = 4.3k_B$. 
Combining these results with the lattice site density, we find an equilibrium vacancy concentration of $C_{V,{\rm Ti}}^{\rm eq}=1.8\times10^{21} {\rm cm}^{-3}{\rm exp}(-1.90[{\rm eV}]/(k_BT))$. 
For the unrelaxed cell, we had previously found with the same method $E_f = 2.05$~eV and $S_f = 8.15k_B$~\cite{Antolin2012fast}. 
Overall, the NN relaxation decreases the formation energy by 0.14~eV. 

Next, we examine the NN-relaxation as a function of temperature and extrapolate from it a zero-temperature displacement which can be used for a fully relaxed surrogate structure with all neighbor shells relaxed. 
Having three temperatures, extrapolation can either be done linearly (dashed line in Fig.~\ref{fig:fig4_NNrelax} a) or with a second-degree polynomial. 
We perform here both and take their average for our zero-temperature structure. 
For a linear extrapolation, we find a zero-temperature displacement of -0.035~{\AA}, for a quadratic one -0.058~{\AA}, with an average of -0.047~{\AA}.
All of these are considerably smaller than the minimum-energy zero-temperature relaxation of -0.101~{\AA}, which is another expression of the mechanical instability of the structure and its strong stabilization by entropy.  
Following the procedure described in Sec.~\ref{subsec:method_relaxation} of interpolating fully relaxed cells of mechanically stable Mo and Ta vacancies with scaled lattice constants, which have NN relaxations of –0.008~{\AA} and –0.082~{\AA} in cells with otherwise clamped other neighbors, we then create a ``relaxed'' $\beta$-Ti cell which is more or less the straight average of the atomic positions in relaxed Mo and Ta cells, scaled to the $\beta$-Ti lattice constant. 
Both ground state structure and saddle point configuration were determined this way. 

The formation energy from this process agrees with the formation energy from the interpolated structure discussed in Sec.~\ref{subsec:method_relaxation} within 0.01~eV, which is less than the uncertainties in our numbers from the different approximations and numerical calculations such as the 0.04~eV from enthalpy discussed in the present section. 
This suggests that relaxation beyond nearest neighbors adds only small changes to the calculated formation energies.

\subsection{\label{subsec:results_Mo_diffusion}Self-Diffusion in Mo}
\begin{figure}
    \centering
    \includegraphics[width=\linewidth]{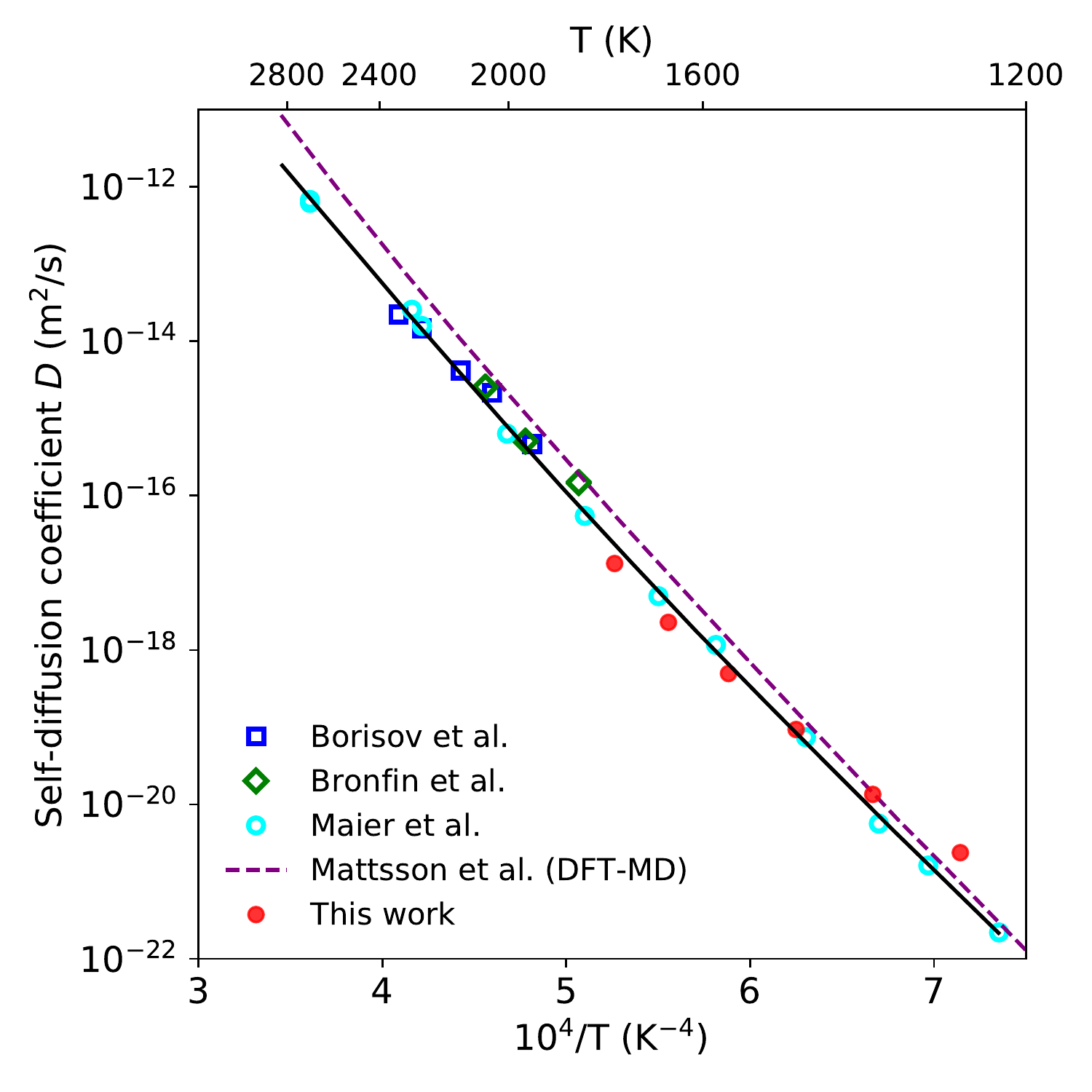}
    \caption{Comparison between calculated self-diffusion coefficient in bcc Mo with experimental data from Refs.~\onlinecite{borisov1959self,Bronfin1960self,maier1979self} and DFT-MD results~\cite{Mattsson2009quantifying}. 
    The solid line represents a parabolic fit of the experimental data by Maier et al.~\cite{maier1979self}. }
    \label{fig:fig6_molly_D}
\end{figure}

Our calculated self-diffusion coefficients for Mo calculated within the quasiharmonic transition-state theory are shown with previous experimental~\cite{borisov1959self,Bronfin1960self,maier1979self} and simulation~\cite{Mattsson2009quantifying} data in Fig.~\ref{fig:fig6_molly_D}. 
Our qhTST calculations with varying lattice constants that include thermal expansion are represented by the six individual points ($1400-1900$~K, with 100~K step). 
We limit our highest calculated temperature to two third of its melting point, which is the commonly accepted applicability range for QHA~\cite{kotelyanskii2004simulation}.

In order to determine the degree of non-linearity in our data, we use Eqs.~\ref{eq:bowing} and ~\ref{eq:non_linearity} and determine a non-linearity parameter of $\lambda = 0.051$. 
While our data thus show good absolute agreement with the experimental results, they have a somewhat higher degree of non-linearity than the experimental values of Maier at al.~\cite{maier1979self}, for which we calculate $\lambda = 0.021$, and also the simulation data of Mattson et al.~\cite{Mattsson2009quantifying} with $\lambda = 0.022$.

In order to explore the origin of the bowing anomaly, we now investigate the temperature dependence of formation and migration enthalpies, as well as of formation entropy and attempt frequency. 
The results are shown in Fig.~\ref{fig:fig8_Hf_Sf}(a), (b) and (c). 
The calculated vacancy formation enthalpy $\Delta H_v$ increases monotonously with temperature from 3.08~eV to 3.17~eV between 1400 and 1900~K (Fig.~\ref{fig:fig8_Hf_Sf}(a)) and is slightly larger than values proposed from analyzing experiments of 3.0~eV [44]. 
The calculated vacancy migration enthalpy $\Delta H_m$ decreases slightly from 1.167~eV to 1.163~eV (Fig.~\ref{fig:fig8_Hf_Sf}(a)), slightly smaller than the reported migration enthalpy of 1.30~eV from experimental analysis~\cite{Ullmaier1991atomic}. 
The calculated attempt frequency $\nu^*$ also shows small temperature dependence and only decreases from $2.3\times10^{12}$~Hz to $2.2\times10^{12}$~Hz as temperature increases from 1400 to 1900~K (Fig.~\ref{fig:fig8_Hf_Sf}(c)).
The temperature dependence of the calculated vacancy formation entropy $\Delta S_f$ on the other hand shows a much more pronounced temperature dependence and is found to be the major reason for non-Arrhenius like diffusion in Mo. 
It decreases from $3.27k_B$ at 1400~K to $2.79k_B$  at 1700~K, before recovering back to $3.22k_B$ at 1900~K, as shown in Fig.~\ref{fig:fig8_Hf_Sf}(b). 
The curved shape of $\Delta S_f (T)$ then directly progresses into the curvature of the calculated self-diffusion coefficients in Fig.~\ref{fig:fig6_molly_D}. 
In order to quantify the effect of the different contributions, we perform two fits of Eq.~\ref{eq:bowing}, first one for calculated diffusion coefficients where the prefactor is fixed to the average value of all calculated prefactors (which results in a value of $D_0^{\rm avg}=3.25\times10^{-6}{\rm m}^2/{\rm s})$, and secondly an analogous set of diffusion coefficients where the activation energy is fixed to the average calculated value of $E_0^{\rm avg}=4.268$~eV. 
For the first fit, we find a non-linearity coefficient of which is 13\% of the overall non-linearity, while for fixed $E_a$, we find $\lambda=0.041$, which is 87\%. 
Thus, the non-linearity in vacancy diffusion in Mo is clearly dominated by the non-linearity of the diffusion prefactor as a consequence of anharmonicity from thermal expansion. 
To investigate the effect further, a similar analysis was made with fixed averaged attempt frequency, resulting in a change of non-linearity by less than 1\%, making the non-linearity in the formation entropy the by far most important factor. 
This conclusion is different from the DFT-MD calculation by Mattsson et al.~\cite{Mattsson2009quantifying}, in which the anomaly was mostly attributed to anharmonicity in the vacancy formation enthalpy $\Delta H_v$. 
However, in their simulations, thermal expansion was not considered and all MD simulations were performed at the zero-temperature volume of vacancy and bulk, which neglected important volume-dependent anharmonic effects from thermal expansion. 
The fundamental proportionality between Grüneisen parameter, the  prime indicator for anharmonicity, and thermal expansion~\cite{sayetat1998easy} as well as our results here are strong indicators that thermal expansion cannot be neglected when trying to understand anharmonic effects.
Our findings are further supported by the resistivity measurements by Schwirtlich and Schultz ~\cite{schwirtlich1980quenching}, which find that the temperature dependence of the formation enthalpy is small and constant to within 0.1 eV between 2000 K and 2600 K, whereas Mattson et al. predict a much larger increase of 0.5 eV for that interval.


\subsection{\label{subsec:results_Ti_diffusion}Self-Diffusion in $\beta$-Ti}
\begin{figure}
    \centering
    \includegraphics[width=\linewidth]{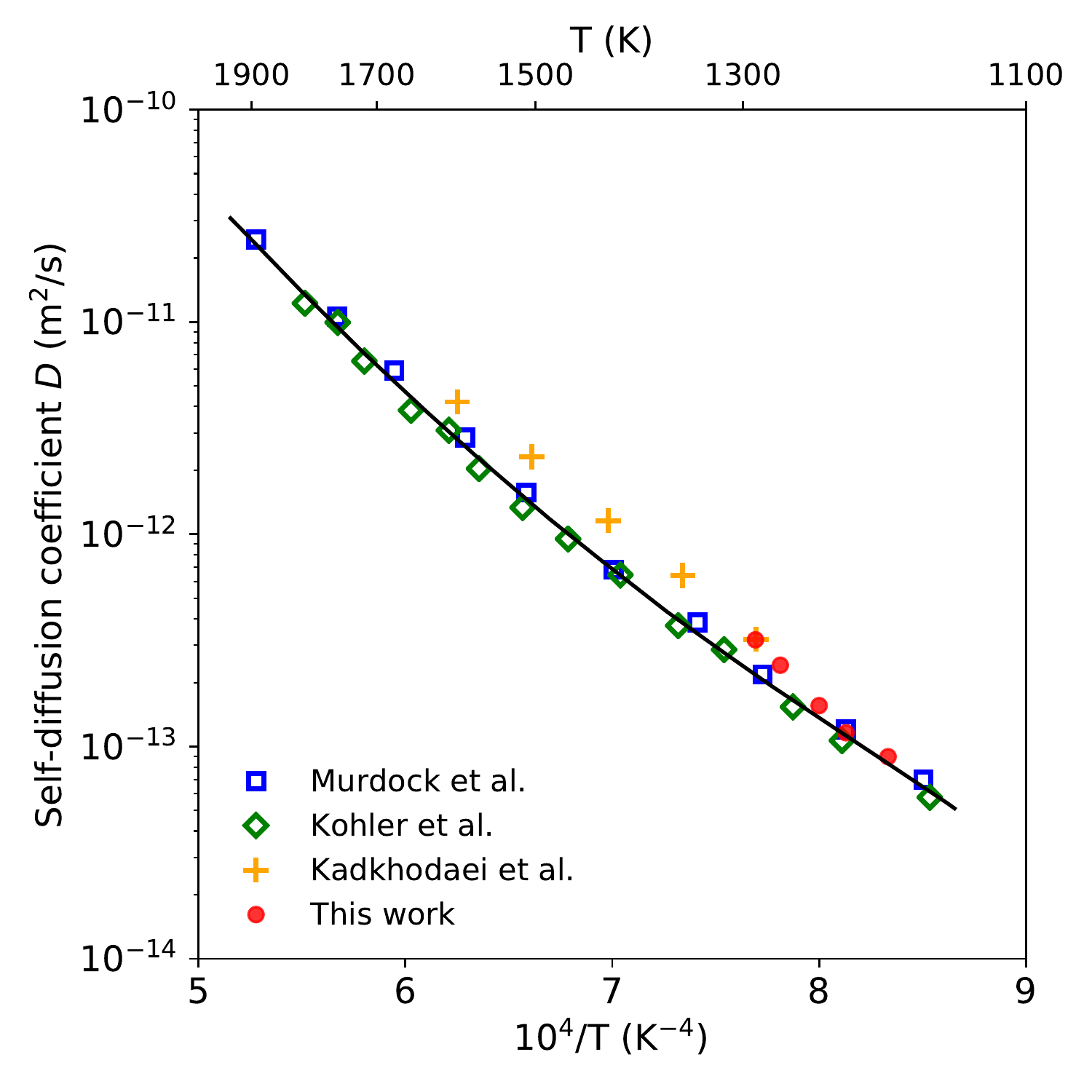}
    \caption{Comparison between calculated self-diffusion coefficients in bcc Ti with experimental data from Ref.~\onlinecite{kohler1987anomalous} and \onlinecite{murdock1964diffusion}. Solid line represents the fitting curve of experimental data.}
    \label{fig:fig7_Ti_D}
\end{figure}

Our calculation results of self-diffusion coefficients in $\beta$-Ti are plotted in Fig.~\ref{fig:fig7_Ti_D} in comparison to experimental data~\cite{kohler1988correlation,murdock1964diffusion}. 
Calculations based on TST and large displacement algorithm are applied at five temperatures: 1200, 1230, 1250, 1280 and 1300~K. 
We choose again the upper limit of 1300~K as two thirds the melting point of Ti (1941~K), while the lower limit of 1200~K is the $\alpha$-to-$\beta$ transition temperature from LDM-DFT calculations [20]. 
Although these choices make the temperature range for the DFT calculations rather small, the anomaly of the self-diffusion coefficient can still be seen in Fig.~\ref{fig:fig7_Ti_D}, and the data are sufficient to extract the non-linearity parameter, which is independent of the temperature interval.

From the LDM-DFT calculations, we find for $\beta$-Ti a non-linearity coefficient of 0.091, about 80\% larger than what we had found for Mo. 
A similarly stronger non-linearity by about a factor of two is also found in experiments, where we fit a non-linearity parameter of $\lambda = 0.046$, which is 120\% larger than the Mo value. 
While thus these trends from experiment are reproduced well by the calculations, and while the values of our diffusion coefficients agree well with experimental values, the calculated non-linearity is considerably stronger than the experimentally observed one, which is true to a similar degree for $\beta$-Ti and for Mo. 

\begin{figure}
    \centering
    \includegraphics[width=\linewidth]{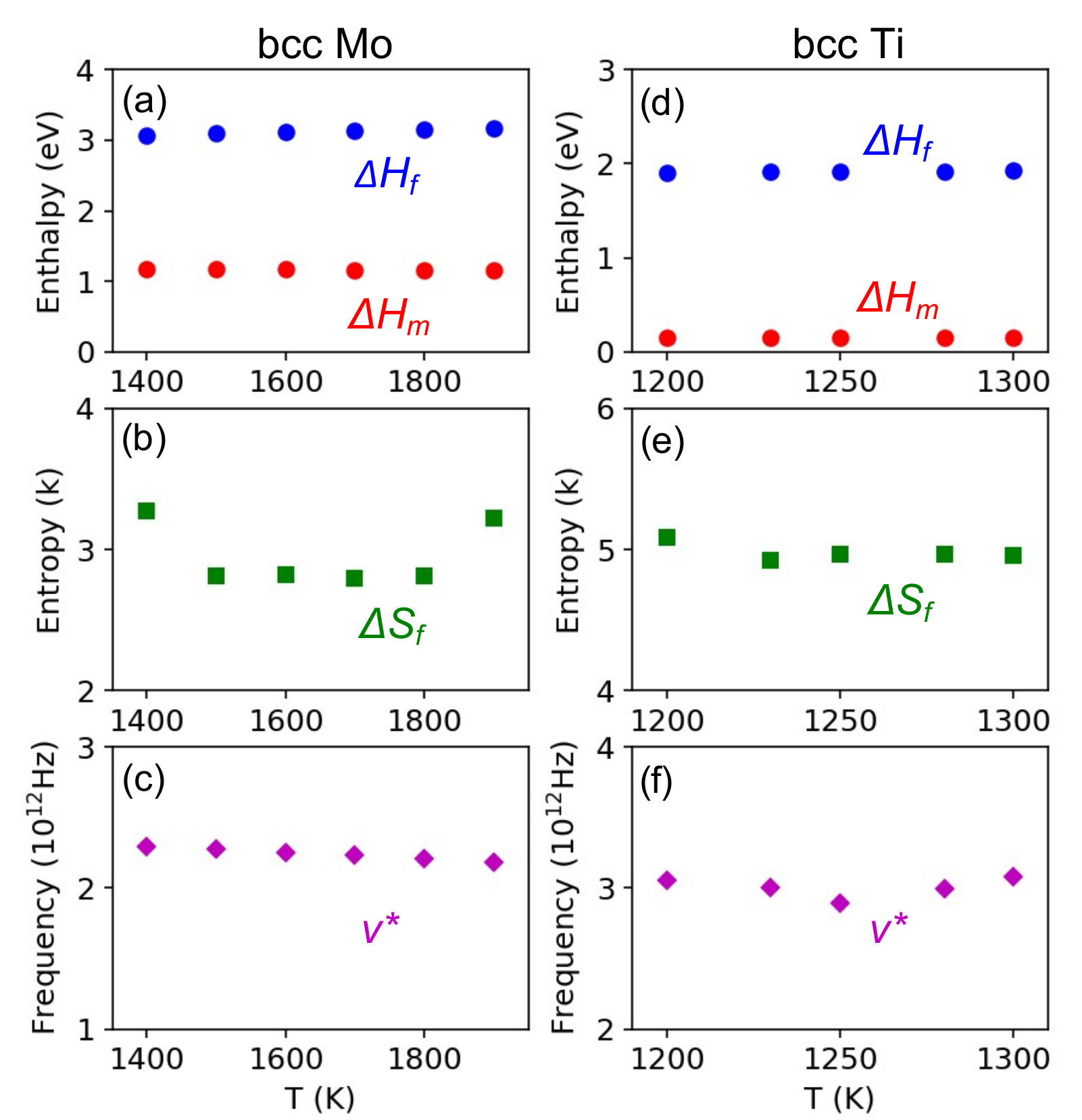}
    \caption{Formation enthalpy $\Delta H_f$, migration enthalpy $\Delta H_m$, formation entropy $\Delta S_f$, and diffusion attempt frequency $\nu^*$ for Mo [(a), (b), (c)] and $\beta$-Ti [(d), (e), (f)], calculated with DFT within quasiharmonic transition state theory.}
    \label{fig:fig8_Hf_Sf}
\end{figure}

In order to explore if the origins of the bowing anomalies in $\beta$-Ti  and Mo differ, we again investigate the temperature dependence of formation and migration enthalpies, as well as of formation entropy and attempt frequency. 
The results are shown in Fig.~\ref{fig:fig8_Hf_Sf}(d), (e) and (f). The vacancy formation enthalpy $\Delta H_f$ increases from 1.91~eV to 1.92~eV at the temperature from 1200~K to 1300~K (Fig.~\ref{fig:fig8_Hf_Sf}(d)), compared with an experimental estimation of $>1.50$~eV~\cite{de1988cohesion}. 
This value from interpolated Mo and Ta zero-temperature cells (Secs.~\ref{subsec:method_relaxation}and \ref{subsec:results_Ti_relaxaion}) agrees well with the results from the nearest-neighbor only free-energy minimization, where the nearly perfect linearity with temperature indicated a very weak temperature dependence of the formation enthalpy (Fig.~\ref{fig:fig8_Hf_Sf}). 
Similarly, the vacancy migration enthalpy $\Delta H_m$ increases only slightly from 0.146~eV to 0.147~eV as temperature increases (Fig.~\ref{fig:fig8_Hf_Sf}(d). 
Indeed, when we fit the calculated diffusion coefficients once again with an averaged prefactor (which has a value of $D_0^{\rm avg}=3.403\times10^{-5}{\rm m}^2/{\rm s}$ for $\beta$-Ti), we only find a non-linearity parameter of $\lambda =0.013$, once again only 15\% of the overall bowing. 
Thus, it is once again the prefactor that we identify as the major source (85\%) of the non-linearity. 
Unlike Mo, however, both vacancy formation entropy and attempt frequency contribute to a more comparable degree to the non-linear Arrhenius curve in $\beta$-Ti. 
Specifically, the vacancy formation entropy $\Delta S_f$ decreases from $5.08k_B$ at 1200~K to a minimum of $4.92k_B$ at 1230~K before increasing back to $4.95k_B$ at 1300~K (Fig.~\ref{fig:fig8_Hf_Sf}(d)) and the attempt frequency $\nu^*$ decreases from $3.1\times10^{12}$~Hz at 1200~K to $2.9\times10^{12}$~Hz at 1250~K, then increases back to $3.1\times10^{12}$~Hz at 1300~K (Fig.~\ref{fig:fig8_Hf_Sf}(f)).
With that, the formation entropy carries 64\% of the non-linearity, leaving the remaining 36\% to the attempt frequency. 
Without the anharmonic attempt frequency, the calculated non-linearity in $\beta$-Ti would only be 25\% larger than that in Mo, and thus the effect of the formation entropy is rather comparable in both materials.

\section{Conclusions and outlook}
In this paper, we have demonstrated that the framework of quasiharmonic transition state theory allows reproducing anomalous self-diffusion in bcc metals for the examples of Mo and $\beta$-Ti by including thermal expansion into the calculations. In order to be able to calculate phonon frequencies needed for attempt frequency and formation entropy in the mechanically unstable $\beta$-Ti, we use the previously proposed large-displacement method~\cite{Antolin2012fast} and introduce two approaches to relax vacancies in mechanically unstable crystals, which is free-energy relaxation of the nearest neighbor shell and for further refinement interpolation of stable structures, where the weight of the structures is determined by comparing nearest-neighbor relaxation displacements. We also propose to use a modified Debye formula with only four parameters to fit thermal expansion over the entire temperature range to be able to predict lattice constants for any temperature easily based on calculated lattice constants for only a few temperatures, since the process to identify the equilibrium lattice constant within the quasiharmonic approximation is somewhat cumbersome. 
We find that phonon frequencies, at the heart of the two quantities that define the diffusion prefactor – vibrational entropy and attempt frequency – change more sensitively with volume changes than formation enthalpy and migration energy and thus dominate the non-linearity in the Arrhenius plots, which is in contrast to previous suggestions.  In order to quantify the degree of non-linearity, we have proposed to use the non-linearity parameter from Ref.~\citenum{emancipator1993quantitative} and find that for both metals studied 85\% of the non-linearity comes from the diffusion prefactor. While these values are similar, we find a considerable difference in the distribution of the non-linearity on formation entropy: While nearly all of the bowing in Mo comes from the formation entropy, the non-linearity in $\beta$-Ti is divided 2/3:1/3 between formation entropy and attempt frequency, respectively, again in contrast to previous suggestions. Our calculated diffusion coefficients agree well with measured values, although we calculate a somewhat larger non-linearity than what we extract from experimental datasets. The proposed methodology is general enough that it also can be applied to other mechanically unstable crystals.

\begin{acknowledgments}
This work is primarily supported by the US National Science Foundation (NSF) under Grant number CMMI-1333999 (ZQC and JCZ), and it is part of an NSF Designing Materials to Revolutionize and Engineer our Future (DMREF) project. Computer calculations were performed at the Ohio Supercomputer Center under grant number PAS0551 and PAS0072. YW and WW acknowledges funding from AFOSR through project number FA9550-19-1-0378.
\end{acknowledgments}


\begin{thebibliography}{57}%
\makeatletter
\providecommand \@ifxundefined [1]{%
 \@ifx{#1\undefined}
}%
\providecommand \@ifnum [1]{%
 \ifnum #1\expandafter \@firstoftwo
 \else \expandafter \@secondoftwo
 \fi
}%
\providecommand \@ifx [1]{%
 \ifx #1\expandafter \@firstoftwo
 \else \expandafter \@secondoftwo
 \fi
}%
\providecommand \natexlab [1]{#1}%
\providecommand \enquote  [1]{``#1''}%
\providecommand \bibnamefont  [1]{#1}%
\providecommand \bibfnamefont [1]{#1}%
\providecommand \citenamefont [1]{#1}%
\providecommand \href@noop [0]{\@secondoftwo}%
\providecommand \href [0]{\begingroup \@sanitize@url \@href}%
\providecommand \@href[1]{\@@startlink{#1}\@@href}%
\providecommand \@@href[1]{\endgroup#1\@@endlink}%
\providecommand \@sanitize@url [0]{\catcode `\\12\catcode `\$12\catcode
  `\&12\catcode `\#12\catcode `\^12\catcode `\_12\catcode `\%12\relax}%
\providecommand \@@startlink[1]{}%
\providecommand \@@endlink[0]{}%
\providecommand \url  [0]{\begingroup\@sanitize@url \@url }%
\providecommand \@url [1]{\endgroup\@href {#1}{\urlprefix }}%
\providecommand \urlprefix  [0]{URL }%
\providecommand \Eprint [0]{\href }%
\providecommand \doibase [0]{http://dx.doi.org/}%
\providecommand \selectlanguage [0]{\@gobble}%
\providecommand \bibinfo  [0]{\@secondoftwo}%
\providecommand \bibfield  [0]{\@secondoftwo}%
\providecommand \translation [1]{[#1]}%
\providecommand \BibitemOpen [0]{}%
\providecommand \bibitemStop [0]{}%
\providecommand \bibitemNoStop [0]{.\EOS\space}%
\providecommand \EOS [0]{\spacefactor3000\relax}%
\providecommand \BibitemShut  [1]{\csname bibitem#1\endcsname}%
\let\auto@bib@innerbib\@empty
\bibitem [{\citenamefont {Neumann}\ and\ \citenamefont
  {Tuijn}(2011)}]{neumann2011self}%
  \BibitemOpen
  \bibfield  {author} {\bibinfo {author} {\bibfnamefont {G.}~\bibnamefont
  {Neumann}}\ and\ \bibinfo {author} {\bibfnamefont {C.}~\bibnamefont
  {Tuijn}},\ }\href@noop {} {\emph {\bibinfo {title} {Self-diffusion and
  impurity diffusion in pure metals: handbook of experimental data}}}\
  (\bibinfo  {publisher} {Elsevier},\ \bibinfo {year} {2011})\BibitemShut
  {NoStop}%
\bibitem [{\citenamefont {Da~Fano}\ and\ \citenamefont
  {Jacucci}(1977)}]{Fano1977vacancy}%
  \BibitemOpen
  \bibfield  {author} {\bibinfo {author} {\bibfnamefont {A.}~\bibnamefont
  {Da~Fano}}\ and\ \bibinfo {author} {\bibfnamefont {G.}~\bibnamefont
  {Jacucci}},\ }\href {\doibase 10.1103/PhysRevLett.39.950} {\bibfield
  {journal} {\bibinfo  {journal} {Phys. Rev. Lett.}\ }\textbf {\bibinfo
  {volume} {39}},\ \bibinfo {pages} {950} (\bibinfo {year} {1977})}\BibitemShut
  {NoStop}%
\bibitem [{\citenamefont {Ait-Salem}\ \emph {et~al.}(1979)\citenamefont
  {Ait-Salem}, \citenamefont {Springer},\ and\ \citenamefont
  {Alefeld}}]{Ait-Salem1979investigation}%
  \BibitemOpen
  \bibfield  {author} {\bibinfo {author} {\bibfnamefont {M.}~\bibnamefont
  {Ait-Salem}}, \bibinfo {author} {\bibfnamefont {A.}~\bibnamefont {Springer},
  \bibfnamefont {T~Heidemann}}, \ and\ \bibinfo {author} {\bibfnamefont
  {B.}~\bibnamefont {Alefeld}},\ }\href
  {https://inis.iaea.org/search/search.aspx?orig_q=RN:10489994} {\bibfield
  {journal} {\bibinfo  {journal} {Philos. Mag. A Phys. Condens. Matter, Struct.
  Defects Mech. Prop.}\ }\textbf {\bibinfo {volume} {39}},\ \bibinfo {pages}
  {797} (\bibinfo {year} {1979})}\BibitemShut {NoStop}%
\bibitem [{\citenamefont {G{\"o}ltz}\ \emph {et~al.}(1980)\citenamefont
  {G{\"o}ltz}, \citenamefont {Heidemann}, \citenamefont {Mehrer}, \citenamefont
  {Seeger},\ and\ \citenamefont {Wolf}}]{goltz1980study}%
  \BibitemOpen
  \bibfield  {author} {\bibinfo {author} {\bibfnamefont {G.}~\bibnamefont
  {G{\"o}ltz}}, \bibinfo {author} {\bibfnamefont {A.}~\bibnamefont
  {Heidemann}}, \bibinfo {author} {\bibfnamefont {H.}~\bibnamefont {Mehrer}},
  \bibinfo {author} {\bibfnamefont {A.}~\bibnamefont {Seeger}}, \ and\ \bibinfo
  {author} {\bibfnamefont {D.}~\bibnamefont {Wolf}},\ }\href@noop {} {\bibfield
   {journal} {\bibinfo  {journal} {Philosophical Magazine A}\ }\textbf
  {\bibinfo {volume} {41}},\ \bibinfo {pages} {723} (\bibinfo {year}
  {1980})}\BibitemShut {NoStop}%
\bibitem [{\citenamefont {Schilling}(1978)}]{schilling1978self}%
  \BibitemOpen
  \bibfield  {author} {\bibinfo {author} {\bibfnamefont {W.}~\bibnamefont
  {Schilling}},\ }\href@noop {} {\bibfield  {journal} {\bibinfo  {journal}
  {Journal of Nuclear Materials}\ }\textbf {\bibinfo {volume} {69}},\ \bibinfo
  {pages} {465} (\bibinfo {year} {1978})}\BibitemShut {NoStop}%
\bibitem [{\citenamefont {Sanchez}\ and\ \citenamefont
  {de~Fontaine}(1975)}]{Sanchez1975model}%
  \BibitemOpen
  \bibfield  {author} {\bibinfo {author} {\bibfnamefont {J.~M.}\ \bibnamefont
  {Sanchez}}\ and\ \bibinfo {author} {\bibfnamefont {D.}~\bibnamefont
  {de~Fontaine}},\ }\href {\doibase 10.1103/PhysRevLett.35.227} {\bibfield
  {journal} {\bibinfo  {journal} {Phys. Rev. Lett.}\ }\textbf {\bibinfo
  {volume} {35}},\ \bibinfo {pages} {227} (\bibinfo {year} {1975})}\BibitemShut
  {NoStop}%
\bibitem [{\citenamefont {Sanchez}\ and\ \citenamefont
  {Fontaine}(1978)}]{Sanchez1978}%
  \BibitemOpen
  \bibfield  {author} {\bibinfo {author} {\bibfnamefont {J.}~\bibnamefont
  {Sanchez}}\ and\ \bibinfo {author} {\bibfnamefont {D.~D.}\ \bibnamefont
  {Fontaine}},\ }\href {\doibase 10.1016/0001-6160(78)90136-0} {\bibfield
  {journal} {\bibinfo  {journal} {Acta Metallurgica}\ }\textbf {\bibinfo
  {volume} {26}},\ \bibinfo {pages} {1083} (\bibinfo {year}
  {1978})}\BibitemShut {NoStop}%
\bibitem [{\citenamefont {Herzig}\ and\ \citenamefont
  {K\"{o}hler}(1987)}]{Herzig1987}%
  \BibitemOpen
  \bibfield  {author} {\bibinfo {author} {\bibfnamefont {C.}~\bibnamefont
  {Herzig}}\ and\ \bibinfo {author} {\bibfnamefont {U.}~\bibnamefont
  {K\"{o}hler}},\ }\href {\doibase 10.4028/www.scientific.net/msf.15-18.301}
  {\bibfield  {journal} {\bibinfo  {journal} {Materials Science Forum}\
  }\textbf {\bibinfo {volume} {15-18}},\ \bibinfo {pages} {301} (\bibinfo
  {year} {1987})}\BibitemShut {NoStop}%
\bibitem [{\citenamefont {Petry}\ \emph {et~al.}(1988)\citenamefont {Petry},
  \citenamefont {Flottmann}, \citenamefont {Heiming}, \citenamefont
  {Trampenau}, \citenamefont {Alba},\ and\ \citenamefont
  {Vogl}}]{Petry1988atomistic}%
  \BibitemOpen
  \bibfield  {author} {\bibinfo {author} {\bibfnamefont {W.}~\bibnamefont
  {Petry}}, \bibinfo {author} {\bibfnamefont {T.}~\bibnamefont {Flottmann}},
  \bibinfo {author} {\bibfnamefont {A.}~\bibnamefont {Heiming}}, \bibinfo
  {author} {\bibfnamefont {J.}~\bibnamefont {Trampenau}}, \bibinfo {author}
  {\bibfnamefont {M.}~\bibnamefont {Alba}}, \ and\ \bibinfo {author}
  {\bibfnamefont {G.}~\bibnamefont {Vogl}},\ }\href {\doibase
  10.1103/PhysRevLett.61.722} {\bibfield  {journal} {\bibinfo  {journal} {Phys.
  Rev. Lett.}\ }\textbf {\bibinfo {volume} {61}},\ \bibinfo {pages} {722}
  (\bibinfo {year} {1988})}\BibitemShut {NoStop}%
\bibitem [{\citenamefont {Vogl}\ \emph {et~al.}(1989)\citenamefont {Vogl},
  \citenamefont {Petry}, \citenamefont {Flottmann},\ and\ \citenamefont
  {Heiming}}]{Vogl1989direct}%
  \BibitemOpen
  \bibfield  {author} {\bibinfo {author} {\bibfnamefont {G.}~\bibnamefont
  {Vogl}}, \bibinfo {author} {\bibfnamefont {W.}~\bibnamefont {Petry}},
  \bibinfo {author} {\bibfnamefont {T.}~\bibnamefont {Flottmann}}, \ and\
  \bibinfo {author} {\bibfnamefont {A.}~\bibnamefont {Heiming}},\ }\href
  {\doibase 10.1103/PhysRevB.39.5025} {\bibfield  {journal} {\bibinfo
  {journal} {Phys. Rev. B}\ }\textbf {\bibinfo {volume} {39}},\ \bibinfo
  {pages} {5025} (\bibinfo {year} {1989})}\BibitemShut {NoStop}%
\bibitem [{\citenamefont {Neumann}\ \emph {et~al.}(1988)\citenamefont
  {Neumann}, \citenamefont {Tuijn}, \citenamefont {de~Vries},\ and\
  \citenamefont {Bakker}}]{Neumann1988}%
  \BibitemOpen
  \bibfield  {author} {\bibinfo {author} {\bibfnamefont {G.}~\bibnamefont
  {Neumann}}, \bibinfo {author} {\bibfnamefont {C.}~\bibnamefont {Tuijn}},
  \bibinfo {author} {\bibfnamefont {G.}~\bibnamefont {de~Vries}}, \ and\
  \bibinfo {author} {\bibfnamefont {H.}~\bibnamefont {Bakker}},\ }\href
  {\doibase 10.1002/pssb.2221490209} {\bibfield  {journal} {\bibinfo  {journal}
  {physica status solidi (b)}\ }\textbf {\bibinfo {volume} {149}},\ \bibinfo
  {pages} {483} (\bibinfo {year} {1988})}\BibitemShut {NoStop}%
\bibitem [{\citenamefont {Rieth}\ and\ \citenamefont
  {Schommers}(2007)}]{rieth2007handbook}%
  \BibitemOpen
  \bibfield  {author} {\bibinfo {author} {\bibfnamefont {M.}~\bibnamefont
  {Rieth}}\ and\ \bibinfo {author} {\bibfnamefont {W.}~\bibnamefont
  {Schommers}},\ }\href@noop {} {\emph {\bibinfo {title} {Handbook Of
  Theoretical And Computational Nanotechnology. Volume 10: Nanodevice Modeling
  And Nanoeletronics}}}\ (\bibinfo  {publisher} {American Scientific
  Publishers},\ \bibinfo {year} {2007})\ pp.\ \bibinfo {pages}
  {137--209}\BibitemShut {NoStop}%
\bibitem [{\citenamefont {Stoddard}\ \emph {et~al.}(2005)\citenamefont
  {Stoddard}, \citenamefont {Pichler}, \citenamefont {Duscher},\ and\
  \citenamefont {Windl}}]{Stoddard20025abinitio}%
  \BibitemOpen
  \bibfield  {author} {\bibinfo {author} {\bibfnamefont {N.}~\bibnamefont
  {Stoddard}}, \bibinfo {author} {\bibfnamefont {P.}~\bibnamefont {Pichler}},
  \bibinfo {author} {\bibfnamefont {G.}~\bibnamefont {Duscher}}, \ and\
  \bibinfo {author} {\bibfnamefont {W.}~\bibnamefont {Windl}},\ }\href
  {\doibase 10.1103/PhysRevLett.95.025901} {\bibfield  {journal} {\bibinfo
  {journal} {Phys. Rev. Lett.}\ }\textbf {\bibinfo {volume} {95}},\ \bibinfo
  {pages} {025901} (\bibinfo {year} {2005})}\BibitemShut {NoStop}%
\bibitem [{\citenamefont {K{\"o}hler}\ and\ \citenamefont
  {Herzig}(1987)}]{kohler1987anomalous}%
  \BibitemOpen
  \bibfield  {author} {\bibinfo {author} {\bibfnamefont {U.}~\bibnamefont
  {K{\"o}hler}}\ and\ \bibinfo {author} {\bibfnamefont {C.}~\bibnamefont
  {Herzig}},\ }\href@noop {} {\bibfield  {journal} {\bibinfo  {journal}
  {physica status solidi (b)}\ }\textbf {\bibinfo {volume} {144}},\ \bibinfo
  {pages} {243} (\bibinfo {year} {1987})}\BibitemShut {NoStop}%
\bibitem [{\citenamefont {K{\"o}hler}\ and\ \citenamefont
  {Herzig}(1988)}]{kohler1988correlation}%
  \BibitemOpen
  \bibfield  {author} {\bibinfo {author} {\bibfnamefont {U.}~\bibnamefont
  {K{\"o}hler}}\ and\ \bibinfo {author} {\bibfnamefont {C.}~\bibnamefont
  {Herzig}},\ }\href@noop {} {\bibfield  {journal} {\bibinfo  {journal}
  {Philosophical Magazine A}\ }\textbf {\bibinfo {volume} {58}},\ \bibinfo
  {pages} {769} (\bibinfo {year} {1988})}\BibitemShut {NoStop}%
\bibitem [{\citenamefont {Smirnov}(2020)}]{Smirnov2020non-arrhenius}%
  \BibitemOpen
  \bibfield  {author} {\bibinfo {author} {\bibfnamefont {G.}~\bibnamefont
  {Smirnov}},\ }\href {\doibase 10.1103/PhysRevB.102.184110} {\bibfield
  {journal} {\bibinfo  {journal} {Phys. Rev. B}\ }\textbf {\bibinfo {volume}
  {102}},\ \bibinfo {pages} {184110} (\bibinfo {year} {2020})}\BibitemShut
  {NoStop}%
\bibitem [{\citenamefont {Kadkhodaei}\ and\ \citenamefont
  {Davariashtiyani}(2020)}]{Kadhodaei2020Phonon}%
  \BibitemOpen
  \bibfield  {author} {\bibinfo {author} {\bibfnamefont {S.}~\bibnamefont
  {Kadkhodaei}}\ and\ \bibinfo {author} {\bibfnamefont {A.}~\bibnamefont
  {Davariashtiyani}},\ }\href {\doibase 10.1103/PhysRevMaterials.4.043802}
  {\bibfield  {journal} {\bibinfo  {journal} {Phys. Rev. Materials}\ }\textbf
  {\bibinfo {volume} {4}},\ \bibinfo {pages} {043802} (\bibinfo {year}
  {2020})}\BibitemShut {NoStop}%
\bibitem [{\citenamefont {Mattsson}\ \emph {et~al.}(2009)\citenamefont
  {Mattsson}, \citenamefont {Sandberg}, \citenamefont {Armiento},\ and\
  \citenamefont {Mattsson}}]{Mattsson2009quantifying}%
  \BibitemOpen
  \bibfield  {author} {\bibinfo {author} {\bibfnamefont {T.~R.}\ \bibnamefont
  {Mattsson}}, \bibinfo {author} {\bibfnamefont {N.}~\bibnamefont {Sandberg}},
  \bibinfo {author} {\bibfnamefont {R.}~\bibnamefont {Armiento}}, \ and\
  \bibinfo {author} {\bibfnamefont {A.~E.}\ \bibnamefont {Mattsson}},\ }\href
  {\doibase 10.1103/PhysRevB.80.224104} {\bibfield  {journal} {\bibinfo
  {journal} {Phys. Rev. B}\ }\textbf {\bibinfo {volume} {80}},\ \bibinfo
  {pages} {224104} (\bibinfo {year} {2009})}\BibitemShut {NoStop}%
\bibitem [{\citenamefont {Sangiovanni}\ \emph {et~al.}(2019)\citenamefont
  {Sangiovanni}, \citenamefont {Klarbring}, \citenamefont {Smirnova},
  \citenamefont {Skripnyak}, \citenamefont {Gambino}, \citenamefont {Mrovec},
  \citenamefont {Simak},\ and\ \citenamefont
  {Abrikosov}}]{Sangiovanni2019superioniclike}%
  \BibitemOpen
  \bibfield  {author} {\bibinfo {author} {\bibfnamefont {D.~G.}\ \bibnamefont
  {Sangiovanni}}, \bibinfo {author} {\bibfnamefont {J.}~\bibnamefont
  {Klarbring}}, \bibinfo {author} {\bibfnamefont {D.}~\bibnamefont {Smirnova}},
  \bibinfo {author} {\bibfnamefont {N.~V.}\ \bibnamefont {Skripnyak}}, \bibinfo
  {author} {\bibfnamefont {D.}~\bibnamefont {Gambino}}, \bibinfo {author}
  {\bibfnamefont {M.}~\bibnamefont {Mrovec}}, \bibinfo {author} {\bibfnamefont
  {S.~I.}\ \bibnamefont {Simak}}, \ and\ \bibinfo {author} {\bibfnamefont
  {I.~A.}\ \bibnamefont {Abrikosov}},\ }\href {\doibase
  10.1103/PhysRevLett.123.105501} {\bibfield  {journal} {\bibinfo  {journal}
  {Phys. Rev. Lett.}\ }\textbf {\bibinfo {volume} {123}},\ \bibinfo {pages}
  {105501} (\bibinfo {year} {2019})}\BibitemShut {NoStop}%
\bibitem [{\citenamefont {Fransson}\ and\ \citenamefont
  {Erhart}(2020)}]{Fransson2020}%
  \BibitemOpen
  \bibfield  {author} {\bibinfo {author} {\bibfnamefont {E.}~\bibnamefont
  {Fransson}}\ and\ \bibinfo {author} {\bibfnamefont {P.}~\bibnamefont
  {Erhart}},\ }\href {\doibase 10.1016/j.actamat.2020.06.040} {\bibfield
  {journal} {\bibinfo  {journal} {Acta Materialia}\ }\textbf {\bibinfo {volume}
  {196}},\ \bibinfo {pages} {770} (\bibinfo {year} {2020})}\BibitemShut
  {NoStop}%
\bibitem [{\citenamefont {Ko}\ \emph {et~al.}(2015)\citenamefont {Ko},
  \citenamefont {Grabowski},\ and\ \citenamefont {Neugebauer}}]{Ko2015}%
  \BibitemOpen
  \bibfield  {author} {\bibinfo {author} {\bibfnamefont {W.-S.}\ \bibnamefont
  {Ko}}, \bibinfo {author} {\bibfnamefont {B.}~\bibnamefont {Grabowski}}, \
  and\ \bibinfo {author} {\bibfnamefont {J.}~\bibnamefont {Neugebauer}},\
  }\href {\doibase 10.1103/PhysRevB.92.134107} {\bibfield  {journal} {\bibinfo
  {journal} {Phys. Rev. B}\ }\textbf {\bibinfo {volume} {92}},\ \bibinfo
  {pages} {134107} (\bibinfo {year} {2015})}\BibitemShut {NoStop}%
\bibitem [{\citenamefont {Hennig}\ \emph {et~al.}(2008)\citenamefont {Hennig},
  \citenamefont {Lenosky}, \citenamefont {Trinkle}, \citenamefont {Rudin},\
  and\ \citenamefont {Wilkins}}]{Hennig2008}%
  \BibitemOpen
  \bibfield  {author} {\bibinfo {author} {\bibfnamefont {R.~G.}\ \bibnamefont
  {Hennig}}, \bibinfo {author} {\bibfnamefont {T.~J.}\ \bibnamefont {Lenosky}},
  \bibinfo {author} {\bibfnamefont {D.~R.}\ \bibnamefont {Trinkle}}, \bibinfo
  {author} {\bibfnamefont {S.~P.}\ \bibnamefont {Rudin}}, \ and\ \bibinfo
  {author} {\bibfnamefont {J.~W.}\ \bibnamefont {Wilkins}},\ }\href {\doibase
  10.1103/PhysRevB.78.054121} {\bibfield  {journal} {\bibinfo  {journal} {Phys.
  Rev. B}\ }\textbf {\bibinfo {volume} {78}},\ \bibinfo {pages} {054121}
  (\bibinfo {year} {2008})}\BibitemShut {NoStop}%
\bibitem [{\citenamefont {Dickel}\ \emph {et~al.}(2018)\citenamefont {Dickel},
  \citenamefont {Barrett}, \citenamefont {Carino}, \citenamefont {Baskes},\
  and\ \citenamefont {Horstemeyer}}]{dickel2018mechanical}%
  \BibitemOpen
  \bibfield  {author} {\bibinfo {author} {\bibfnamefont {D.}~\bibnamefont
  {Dickel}}, \bibinfo {author} {\bibfnamefont {C.~D.}\ \bibnamefont {Barrett}},
  \bibinfo {author} {\bibfnamefont {R.~L.}\ \bibnamefont {Carino}}, \bibinfo
  {author} {\bibfnamefont {M.~I.}\ \bibnamefont {Baskes}}, \ and\ \bibinfo
  {author} {\bibfnamefont {M.~F.}\ \bibnamefont {Horstemeyer}},\ }\href
  {\doibase https://iopscience.iop.org/article/10.1088/1361-651X/aac95d/meta}
  {\bibfield  {journal} {\bibinfo  {journal} {Modelling and Simulation in
  Materials Science and Engineering}\ }\textbf {\bibinfo {volume} {26}},\
  \bibinfo {pages} {065002} (\bibinfo {year} {2018})}\BibitemShut {NoStop}%
\bibitem [{\citenamefont {Windl}\ \emph {et~al.}(1999)\citenamefont {Windl},
  \citenamefont {Bunea}, \citenamefont {Stumpf}, \citenamefont {Dunham},\ and\
  \citenamefont {Masquelier}}]{Windl1999}%
  \BibitemOpen
  \bibfield  {author} {\bibinfo {author} {\bibfnamefont {W.}~\bibnamefont
  {Windl}}, \bibinfo {author} {\bibfnamefont {M.~M.}\ \bibnamefont {Bunea}},
  \bibinfo {author} {\bibfnamefont {R.}~\bibnamefont {Stumpf}}, \bibinfo
  {author} {\bibfnamefont {S.~T.}\ \bibnamefont {Dunham}}, \ and\ \bibinfo
  {author} {\bibfnamefont {M.~P.}\ \bibnamefont {Masquelier}},\ }\href
  {\doibase 10.1103/PhysRevLett.83.4345} {\bibfield  {journal} {\bibinfo
  {journal} {Phys. Rev. Lett.}\ }\textbf {\bibinfo {volume} {83}},\ \bibinfo
  {pages} {4345} (\bibinfo {year} {1999})}\BibitemShut {NoStop}%
\bibitem [{\citenamefont {Wu}\ \emph {et~al.}(2016)\citenamefont {Wu},
  \citenamefont {Mayeshiba},\ and\ \citenamefont {Morgan}}]{wu2016high}%
  \BibitemOpen
  \bibfield  {author} {\bibinfo {author} {\bibfnamefont {H.}~\bibnamefont
  {Wu}}, \bibinfo {author} {\bibfnamefont {T.}~\bibnamefont {Mayeshiba}}, \
  and\ \bibinfo {author} {\bibfnamefont {D.}~\bibnamefont {Morgan}},\
  }\href@noop {} {\bibfield  {journal} {\bibinfo  {journal} {Scientific data}\
  }\textbf {\bibinfo {volume} {3}},\ \bibinfo {pages} {1} (\bibinfo {year}
  {2016})}\BibitemShut {NoStop}%
\bibitem [{\citenamefont {Angsten}\ \emph {et~al.}(2014)\citenamefont
  {Angsten}, \citenamefont {Mayeshiba}, \citenamefont {Wu},\ and\ \citenamefont
  {Morgan}}]{angsten2014elemental}%
  \BibitemOpen
  \bibfield  {author} {\bibinfo {author} {\bibfnamefont {T.}~\bibnamefont
  {Angsten}}, \bibinfo {author} {\bibfnamefont {T.}~\bibnamefont {Mayeshiba}},
  \bibinfo {author} {\bibfnamefont {H.}~\bibnamefont {Wu}}, \ and\ \bibinfo
  {author} {\bibfnamefont {D.}~\bibnamefont {Morgan}},\ }\href@noop {}
  {\bibfield  {journal} {\bibinfo  {journal} {New Journal of Physics}\ }\textbf
  {\bibinfo {volume} {16}},\ \bibinfo {pages} {015018} (\bibinfo {year}
  {2014})}\BibitemShut {NoStop}%
\bibitem [{\citenamefont {Mantina}\ \emph {et~al.}(2008)\citenamefont
  {Mantina}, \citenamefont {Wang}, \citenamefont {Arroyave}, \citenamefont
  {Chen}, \citenamefont {Liu},\ and\ \citenamefont
  {Wolverton}}]{Mantina2008first}%
  \BibitemOpen
  \bibfield  {author} {\bibinfo {author} {\bibfnamefont {M.}~\bibnamefont
  {Mantina}}, \bibinfo {author} {\bibfnamefont {Y.}~\bibnamefont {Wang}},
  \bibinfo {author} {\bibfnamefont {R.}~\bibnamefont {Arroyave}}, \bibinfo
  {author} {\bibfnamefont {L.~Q.}\ \bibnamefont {Chen}}, \bibinfo {author}
  {\bibfnamefont {Z.~K.}\ \bibnamefont {Liu}}, \ and\ \bibinfo {author}
  {\bibfnamefont {C.}~\bibnamefont {Wolverton}},\ }\href {\doibase
  10.1103/PhysRevLett.100.215901} {\bibfield  {journal} {\bibinfo  {journal}
  {Phys. Rev. Lett.}\ }\textbf {\bibinfo {volume} {100}},\ \bibinfo {pages}
  {215901} (\bibinfo {year} {2008})}\BibitemShut {NoStop}%
\bibitem [{\citenamefont {Mantina}\ \emph {et~al.}(2009)\citenamefont
  {Mantina}, \citenamefont {Wang}, \citenamefont {Chen}, \citenamefont {Liu},\
  and\ \citenamefont {Wolverton}}]{mantina2009first}%
  \BibitemOpen
  \bibfield  {author} {\bibinfo {author} {\bibfnamefont {M.}~\bibnamefont
  {Mantina}}, \bibinfo {author} {\bibfnamefont {Y.}~\bibnamefont {Wang}},
  \bibinfo {author} {\bibfnamefont {L.}~\bibnamefont {Chen}}, \bibinfo {author}
  {\bibfnamefont {Z.}~\bibnamefont {Liu}}, \ and\ \bibinfo {author}
  {\bibfnamefont {C.}~\bibnamefont {Wolverton}},\ }\href@noop {} {\bibfield
  {journal} {\bibinfo  {journal} {Acta Materialia}\ }\textbf {\bibinfo {volume}
  {57}},\ \bibinfo {pages} {4102} (\bibinfo {year} {2009})}\BibitemShut
  {NoStop}%
\bibitem [{\citenamefont {Wagner}\ and\ \citenamefont
  {Windl}(2008)}]{wagner2008lattice}%
  \BibitemOpen
  \bibfield  {author} {\bibinfo {author} {\bibfnamefont {M.-X.}\ \bibnamefont
  {Wagner}}\ and\ \bibinfo {author} {\bibfnamefont {W.}~\bibnamefont {Windl}},\
  }\href@noop {} {\bibfield  {journal} {\bibinfo  {journal} {Acta Materialia}\
  }\textbf {\bibinfo {volume} {56}},\ \bibinfo {pages} {6232} (\bibinfo {year}
  {2008})}\BibitemShut {NoStop}%
\bibitem [{\citenamefont {Antolin}\ \emph {et~al.}(2012)\citenamefont
  {Antolin}, \citenamefont {Restrepo},\ and\ \citenamefont
  {Windl}}]{Antolin2012fast}%
  \BibitemOpen
  \bibfield  {author} {\bibinfo {author} {\bibfnamefont {N.}~\bibnamefont
  {Antolin}}, \bibinfo {author} {\bibfnamefont {O.~D.}\ \bibnamefont
  {Restrepo}}, \ and\ \bibinfo {author} {\bibfnamefont {W.}~\bibnamefont
  {Windl}},\ }\href {\doibase 10.1103/PhysRevB.86.054119} {\bibfield  {journal}
  {\bibinfo  {journal} {Phys. Rev. B}\ }\textbf {\bibinfo {volume} {86}},\
  \bibinfo {pages} {054119} (\bibinfo {year} {2012})}\BibitemShut {NoStop}%
\bibitem [{\citenamefont {Hellman}\ \emph {et~al.}(2011)\citenamefont
  {Hellman}, \citenamefont {Abrikosov},\ and\ \citenamefont
  {Simak}}]{Hellman2011lattice}%
  \BibitemOpen
  \bibfield  {author} {\bibinfo {author} {\bibfnamefont {O.}~\bibnamefont
  {Hellman}}, \bibinfo {author} {\bibfnamefont {I.~A.}\ \bibnamefont
  {Abrikosov}}, \ and\ \bibinfo {author} {\bibfnamefont {S.~I.}\ \bibnamefont
  {Simak}},\ }\href {\doibase 10.1103/PhysRevB.84.180301} {\bibfield  {journal}
  {\bibinfo  {journal} {Phys. Rev. B}\ }\textbf {\bibinfo {volume} {84}},\
  \bibinfo {pages} {180301} (\bibinfo {year} {2011})}\BibitemShut {NoStop}%
\bibitem [{\citenamefont {Souvatzis}\ \emph {et~al.}(2009)\citenamefont
  {Souvatzis}, \citenamefont {Eriksson}, \citenamefont {Katsnelson},\ and\
  \citenamefont {Rudin}}]{souvatzis2009self}%
  \BibitemOpen
  \bibfield  {author} {\bibinfo {author} {\bibfnamefont {P.}~\bibnamefont
  {Souvatzis}}, \bibinfo {author} {\bibfnamefont {O.}~\bibnamefont {Eriksson}},
  \bibinfo {author} {\bibfnamefont {M.}~\bibnamefont {Katsnelson}}, \ and\
  \bibinfo {author} {\bibfnamefont {S.}~\bibnamefont {Rudin}},\ }\href@noop {}
  {\bibfield  {journal} {\bibinfo  {journal} {Computational materials science}\
  }\textbf {\bibinfo {volume} {44}},\ \bibinfo {pages} {888} (\bibinfo {year}
  {2009})}\BibitemShut {NoStop}%
\bibitem [{\citenamefont {Montet}(1973)}]{Montet1973integral}%
  \BibitemOpen
  \bibfield  {author} {\bibinfo {author} {\bibfnamefont {G.~L.}\ \bibnamefont
  {Montet}},\ }\href {\doibase 10.1103/PhysRevB.7.650} {\bibfield  {journal}
  {\bibinfo  {journal} {Phys. Rev. B}\ }\textbf {\bibinfo {volume} {7}},\
  \bibinfo {pages} {650} (\bibinfo {year} {1973})}\BibitemShut {NoStop}%
\bibitem [{\citenamefont {Compaan}\ and\ \citenamefont
  {Haven}(1956)}]{compaan1956correlation}%
  \BibitemOpen
  \bibfield  {author} {\bibinfo {author} {\bibfnamefont {K.}~\bibnamefont
  {Compaan}}\ and\ \bibinfo {author} {\bibfnamefont {Y.}~\bibnamefont
  {Haven}},\ }\href@noop {} {\bibfield  {journal} {\bibinfo  {journal}
  {Transactions of the Faraday Society}\ }\textbf {\bibinfo {volume} {52}},\
  \bibinfo {pages} {786} (\bibinfo {year} {1956})}\BibitemShut {NoStop}%
\bibitem [{\citenamefont {Vineyard}(1957)}]{vineyard1957frequency}%
  \BibitemOpen
  \bibfield  {author} {\bibinfo {author} {\bibfnamefont {G.~H.}\ \bibnamefont
  {Vineyard}},\ }\href@noop {} {\bibfield  {journal} {\bibinfo  {journal}
  {Journal of Physics and Chemistry of Solids}\ }\textbf {\bibinfo {volume}
  {3}},\ \bibinfo {pages} {121} (\bibinfo {year} {1957})}\BibitemShut {NoStop}%
\bibitem [{\citenamefont {Baroni}\ \emph {et~al.}(2001)\citenamefont {Baroni},
  \citenamefont {de~Gironcoli}, \citenamefont {Dal~Corso},\ and\ \citenamefont
  {Giannozzi}}]{Baroni2001RMP}%
  \BibitemOpen
  \bibfield  {author} {\bibinfo {author} {\bibfnamefont {S.}~\bibnamefont
  {Baroni}}, \bibinfo {author} {\bibfnamefont {S.}~\bibnamefont
  {de~Gironcoli}}, \bibinfo {author} {\bibfnamefont {A.}~\bibnamefont
  {Dal~Corso}}, \ and\ \bibinfo {author} {\bibfnamefont {P.}~\bibnamefont
  {Giannozzi}},\ }\href {\doibase 10.1103/RevModPhys.73.515} {\bibfield
  {journal} {\bibinfo  {journal} {Rev. Mod. Phys.}\ }\textbf {\bibinfo {volume}
  {73}},\ \bibinfo {pages} {515} (\bibinfo {year} {2001})}\BibitemShut
  {NoStop}%
\bibitem [{\citenamefont {Luo}\ and\ \citenamefont
  {Windl}(2009)}]{luo2009first}%
  \BibitemOpen
  \bibfield  {author} {\bibinfo {author} {\bibfnamefont {W.}~\bibnamefont
  {Luo}}\ and\ \bibinfo {author} {\bibfnamefont {W.}~\bibnamefont {Windl}},\
  }\href@noop {} {\bibfield  {journal} {\bibinfo  {journal} {Carbon}\ }\textbf
  {\bibinfo {volume} {47}},\ \bibinfo {pages} {367} (\bibinfo {year}
  {2009})}\BibitemShut {NoStop}%
\bibitem [{\citenamefont {Henkelman}\ \emph {et~al.}(2000)\citenamefont
  {Henkelman}, \citenamefont {Uberuaga},\ and\ \citenamefont
  {J{\'o}nsson}}]{henkelman2000climbing}%
  \BibitemOpen
  \bibfield  {author} {\bibinfo {author} {\bibfnamefont {G.}~\bibnamefont
  {Henkelman}}, \bibinfo {author} {\bibfnamefont {B.~P.}\ \bibnamefont
  {Uberuaga}}, \ and\ \bibinfo {author} {\bibfnamefont {H.}~\bibnamefont
  {J{\'o}nsson}},\ }\href@noop {} {\bibfield  {journal} {\bibinfo  {journal}
  {The Journal of chemical physics}\ }\textbf {\bibinfo {volume} {113}},\
  \bibinfo {pages} {9901} (\bibinfo {year} {2000})}\BibitemShut {NoStop}%
\bibitem [{\citenamefont {Sayetat}\ \emph {et~al.}(1998)\citenamefont
  {Sayetat}, \citenamefont {Fertey},\ and\ \citenamefont
  {Kessler}}]{sayetat1998easy}%
  \BibitemOpen
  \bibfield  {author} {\bibinfo {author} {\bibfnamefont {F.}~\bibnamefont
  {Sayetat}}, \bibinfo {author} {\bibfnamefont {P.}~\bibnamefont {Fertey}}, \
  and\ \bibinfo {author} {\bibfnamefont {M.}~\bibnamefont {Kessler}},\
  }\href@noop {} {\bibfield  {journal} {\bibinfo  {journal} {Journal of applied
  crystallography}\ }\textbf {\bibinfo {volume} {31}},\ \bibinfo {pages} {121}
  (\bibinfo {year} {1998})}\BibitemShut {NoStop}%
\bibitem [{\citenamefont {Thacher~Jr}(1960)}]{thacher1960rational}%
  \BibitemOpen
  \bibfield  {author} {\bibinfo {author} {\bibfnamefont {H.~C.}\ \bibnamefont
  {Thacher~Jr}},\ }\href@noop {} {\bibfield  {journal} {\bibinfo  {journal}
  {The Journal of Chemical Physics}\ }\textbf {\bibinfo {volume} {32}},\
  \bibinfo {pages} {638} (\bibinfo {year} {1960})}\BibitemShut {NoStop}%
\bibitem [{\citenamefont {Dickinson}\ and\ \citenamefont
  {Armstrong}(1967)}]{dickinson1967temperature}%
  \BibitemOpen
  \bibfield  {author} {\bibinfo {author} {\bibfnamefont {J.}~\bibnamefont
  {Dickinson}}\ and\ \bibinfo {author} {\bibfnamefont {P.}~\bibnamefont
  {Armstrong}},\ }\href@noop {} {\bibfield  {journal} {\bibinfo  {journal}
  {Journal of applied Physics}\ }\textbf {\bibinfo {volume} {38}},\ \bibinfo
  {pages} {602} (\bibinfo {year} {1967})}\BibitemShut {NoStop}%
\bibitem [{\citenamefont {Windl}\ \emph {et~al.}(1998)\citenamefont {Windl},
  \citenamefont {Sankey},\ and\ \citenamefont {Men\'endez}}]{Windl1998theory}%
  \BibitemOpen
  \bibfield  {author} {\bibinfo {author} {\bibfnamefont {W.}~\bibnamefont
  {Windl}}, \bibinfo {author} {\bibfnamefont {O.~F.}\ \bibnamefont {Sankey}}, \
  and\ \bibinfo {author} {\bibfnamefont {J.}~\bibnamefont {Men\'endez}},\
  }\href {\doibase 10.1103/PhysRevB.57.2431} {\bibfield  {journal} {\bibinfo
  {journal} {Phys. Rev. B}\ }\textbf {\bibinfo {volume} {57}},\ \bibinfo
  {pages} {2431} (\bibinfo {year} {1998})}\BibitemShut {NoStop}%
\bibitem [{\citenamefont {Emancipator}\ and\ \citenamefont
  {Kroll}(1993)}]{emancipator1993quantitative}%
  \BibitemOpen
  \bibfield  {author} {\bibinfo {author} {\bibfnamefont {K.}~\bibnamefont
  {Emancipator}}\ and\ \bibinfo {author} {\bibfnamefont {M.~H.}\ \bibnamefont
  {Kroll}},\ }\href@noop {} {\bibfield  {journal} {\bibinfo  {journal}
  {Clinical chemistry}\ }\textbf {\bibinfo {volume} {39}},\ \bibinfo {pages}
  {766} (\bibinfo {year} {1993})}\BibitemShut {NoStop}%
\bibitem [{\citenamefont {Kresse}\ and\ \citenamefont
  {Joubert}(1999)}]{kresse1999ultrasoft}%
  \BibitemOpen
  \bibfield  {author} {\bibinfo {author} {\bibfnamefont {G.}~\bibnamefont
  {Kresse}}\ and\ \bibinfo {author} {\bibfnamefont {D.}~\bibnamefont
  {Joubert}},\ }\href@noop {} {\bibfield  {journal} {\bibinfo  {journal}
  {Physical review b}\ }\textbf {\bibinfo {volume} {59}},\ \bibinfo {pages}
  {1758} (\bibinfo {year} {1999})}\BibitemShut {NoStop}%
\bibitem [{\citenamefont {Kresse}\ and\ \citenamefont
  {Hafner}(1993)}]{kresse1993ab}%
  \BibitemOpen
  \bibfield  {author} {\bibinfo {author} {\bibfnamefont {G.}~\bibnamefont
  {Kresse}}\ and\ \bibinfo {author} {\bibfnamefont {J.}~\bibnamefont
  {Hafner}},\ }\href@noop {} {\bibfield  {journal} {\bibinfo  {journal}
  {Physical Review B}\ }\textbf {\bibinfo {volume} {47}},\ \bibinfo {pages}
  {558} (\bibinfo {year} {1993})}\BibitemShut {NoStop}%
\bibitem [{\citenamefont {Perdew}\ \emph {et~al.}(1996)\citenamefont {Perdew},
  \citenamefont {Burke},\ and\ \citenamefont
  {Ernzerhof}}]{Perdew1996generalized}%
  \BibitemOpen
  \bibfield  {author} {\bibinfo {author} {\bibfnamefont {J.~P.}\ \bibnamefont
  {Perdew}}, \bibinfo {author} {\bibfnamefont {K.}~\bibnamefont {Burke}}, \
  and\ \bibinfo {author} {\bibfnamefont {M.}~\bibnamefont {Ernzerhof}},\ }\href
  {\doibase 10.1103/PhysRevLett.77.3865} {\bibfield  {journal} {\bibinfo
  {journal} {Phys. Rev. Lett.}\ }\textbf {\bibinfo {volume} {77}},\ \bibinfo
  {pages} {3865} (\bibinfo {year} {1996})}\BibitemShut {NoStop}%
\bibitem [{\citenamefont {Wang}\ and\ \citenamefont
  {Reeber}(1998)}]{wang1998role}%
  \BibitemOpen
  \bibfield  {author} {\bibinfo {author} {\bibfnamefont {K.}~\bibnamefont
  {Wang}}\ and\ \bibinfo {author} {\bibfnamefont {R.~R.}\ \bibnamefont
  {Reeber}},\ }\href@noop {} {\bibfield  {journal} {\bibinfo  {journal}
  {Materials Science and Engineering: R: Reports}\ }\textbf {\bibinfo {volume}
  {23}},\ \bibinfo {pages} {101} (\bibinfo {year} {1998})}\BibitemShut
  {NoStop}%
\bibitem [{\citenamefont {Touloukian}\ \emph {et~al.}(1975)\citenamefont
  {Touloukian}, \citenamefont {Kirby}, \citenamefont {Taylor},\ and\
  \citenamefont {Desai}}]{touloukian1975thermophysical}%
  \BibitemOpen
  \bibfield  {author} {\bibinfo {author} {\bibfnamefont {Y.~S.}\ \bibnamefont
  {Touloukian}}, \bibinfo {author} {\bibfnamefont {R.}~\bibnamefont {Kirby}},
  \bibinfo {author} {\bibfnamefont {R.}~\bibnamefont {Taylor}}, \ and\ \bibinfo
  {author} {\bibfnamefont {P.}~\bibnamefont {Desai}},\ }\href@noop {} {\emph
  {\bibinfo {title} {Thermophysical properties of matter-the tprc data series.
  volume 12. thermal expansion metallic elements and alloys}}},\ \bibinfo
  {type} {Tech. Rep.}\ (\bibinfo  {institution} {THERMOPHYSICAL AND ELECTRONIC
  PROPERTIES INFORMATION ANALYSIS CENTER},\ \bibinfo {year} {1975})\BibitemShut
  {NoStop}%
\bibitem [{\citenamefont {Murnaghan}(1944)}]{Murnaghan1944}%
  \BibitemOpen
  \bibfield  {author} {\bibinfo {author} {\bibfnamefont {F.~D.}\ \bibnamefont
  {Murnaghan}},\ }\href {\doibase 10.1073/pnas.30.9.244} {\bibfield  {journal}
  {\bibinfo  {journal} {Proceedings of the National Academy of Sciences}\
  }\textbf {\bibinfo {volume} {30}},\ \bibinfo {pages} {244} (\bibinfo {year}
  {1944})}\BibitemShut {NoStop}%
\bibitem [{\citenamefont {Borisov}\ \emph {et~al.}(1959)\citenamefont
  {Borisov}, \citenamefont {Gruzin}, \citenamefont {Pavliniv},\ and\
  \citenamefont {Fedorov}}]{borisov1959self}%
  \BibitemOpen
  \bibfield  {author} {\bibinfo {author} {\bibfnamefont {Y.}~\bibnamefont
  {Borisov}}, \bibinfo {author} {\bibfnamefont {P.}~\bibnamefont {Gruzin}},
  \bibinfo {author} {\bibfnamefont {L.}~\bibnamefont {Pavliniv}}, \ and\
  \bibinfo {author} {\bibfnamefont {G.}~\bibnamefont {Fedorov}},\ }\href@noop
  {} {\bibfield  {journal} {\bibinfo  {journal} {Metall. Pure Met. Sci}\
  }\textbf {\bibinfo {volume} {1}},\ \bibinfo {pages} {213} (\bibinfo {year}
  {1959})}\BibitemShut {NoStop}%
\bibitem [{\citenamefont {Bronfin}\ \emph {et~al.}(1960)\citenamefont
  {Bronfin}, \citenamefont {Bokshtein},\ and\ \citenamefont
  {Zhukhovistsky}}]{Bronfin1960self}%
  \BibitemOpen
  \bibfield  {author} {\bibinfo {author} {\bibfnamefont {M.}~\bibnamefont
  {Bronfin}}, \bibinfo {author} {\bibfnamefont {S.}~\bibnamefont {Bokshtein}},
  \ and\ \bibinfo {author} {\bibfnamefont {A.}~\bibnamefont {Zhukhovistsky}},\
  }\href@noop {} {\bibfield  {journal} {\bibinfo  {journal} {Zavod. Lab.}\
  }\textbf {\bibinfo {volume} {26}},\ \bibinfo {pages} {828} (\bibinfo {year}
  {1960})}\BibitemShut {NoStop}%
\bibitem [{\citenamefont {Maier}\ \emph {et~al.}(1979)\citenamefont {Maier},
  \citenamefont {Mehrer},\ and\ \citenamefont {Rein}}]{maier1979self}%
  \BibitemOpen
  \bibfield  {author} {\bibinfo {author} {\bibfnamefont {K.}~\bibnamefont
  {Maier}}, \bibinfo {author} {\bibfnamefont {H.}~\bibnamefont {Mehrer}}, \
  and\ \bibinfo {author} {\bibfnamefont {G.}~\bibnamefont {Rein}},\ }\href@noop
  {} {\bibfield  {journal} {\bibinfo  {journal} {Z. Metallkd.}\ }\textbf
  {\bibinfo {volume} {70}},\ \bibinfo {pages} {271} (\bibinfo {year}
  {1979})}\BibitemShut {NoStop}%
\bibitem [{\citenamefont {Kotelyanskii}\ and\ \citenamefont
  {Theodorou}(2004)}]{kotelyanskii2004simulation}%
  \BibitemOpen
  \bibfield  {author} {\bibinfo {author} {\bibfnamefont {M.}~\bibnamefont
  {Kotelyanskii}}\ and\ \bibinfo {author} {\bibfnamefont {D.~N.}\ \bibnamefont
  {Theodorou}},\ }\href@noop {} {\emph {\bibinfo {title} {Simulation methods
  for polymers}}}\ (\bibinfo  {publisher} {CRC Press},\ \bibinfo {year}
  {2004})\BibitemShut {NoStop}%
\bibitem [{\citenamefont {Ullmaier}(1991)}]{Ullmaier1991atomic}%
  \BibitemOpen
  \bibinfo {editor} {\bibfnamefont {H.}~\bibnamefont {Ullmaier}},\ ed.,\ \href
  {\doibase 10.1007/b37800} {\emph {\bibinfo {title} {Atomic Defects in
  Metals}}}\ (\bibinfo  {publisher} {Springer-Verlag},\ \bibinfo {year}
  {1991})\BibitemShut {NoStop}%
\bibitem [{\citenamefont {Schwirtlich}\ and\ \citenamefont
  {Schultz}(1980)}]{schwirtlich1980quenching}%
  \BibitemOpen
  \bibfield  {author} {\bibinfo {author} {\bibfnamefont {I.}~\bibnamefont
  {Schwirtlich}}\ and\ \bibinfo {author} {\bibfnamefont {H.}~\bibnamefont
  {Schultz}},\ }\href
  {https://www.tandfonline.com/doi/abs/10.1080/01418618008241839} {\bibfield
  {journal} {\bibinfo  {journal} {Philosophical Magazine A}\ }\textbf {\bibinfo
  {volume} {42}},\ \bibinfo {pages} {601} (\bibinfo {year} {1980})}\BibitemShut
  {NoStop}%
\bibitem [{\citenamefont {Murdock}\ \emph {et~al.}(1964)\citenamefont
  {Murdock}, \citenamefont {Lundy},\ and\ \citenamefont
  {Stansbury}}]{murdock1964diffusion}%
  \BibitemOpen
  \bibfield  {author} {\bibinfo {author} {\bibfnamefont {J.}~\bibnamefont
  {Murdock}}, \bibinfo {author} {\bibfnamefont {T.}~\bibnamefont {Lundy}}, \
  and\ \bibinfo {author} {\bibfnamefont {E.}~\bibnamefont {Stansbury}},\
  }\href@noop {} {\bibfield  {journal} {\bibinfo  {journal} {Acta
  metallurgica}\ }\textbf {\bibinfo {volume} {12}},\ \bibinfo {pages} {1033}
  (\bibinfo {year} {1964})}\BibitemShut {NoStop}%
\bibitem [{\citenamefont {De~Boer}\ \emph {et~al.}(1988)\citenamefont
  {De~Boer}, \citenamefont {Mattens}, \citenamefont {Boom}, \citenamefont
  {Miedema},\ and\ \citenamefont {Niessen}}]{de1988cohesion}%
  \BibitemOpen
  \bibfield  {author} {\bibinfo {author} {\bibfnamefont {F.~R.}\ \bibnamefont
  {De~Boer}}, \bibinfo {author} {\bibfnamefont {W.}~\bibnamefont {Mattens}},
  \bibinfo {author} {\bibfnamefont {R.}~\bibnamefont {Boom}}, \bibinfo {author}
  {\bibfnamefont {A.}~\bibnamefont {Miedema}}, \ and\ \bibinfo {author}
  {\bibfnamefont {A.}~\bibnamefont {Niessen}},\ }\href@noop {} {\emph {\bibinfo
  {title} {Cohesion in metals. Transition metal alloys}}}\ (\bibinfo {year}
  {1988})\BibitemShut {NoStop}%
\end{thebibliography}
%

\end{document}